\documentclass[12pt]{amsart}
\usepackage{amsmath}
\usepackage{amsxtra}
\usepackage{amscd}
\usepackage{amsthm}
\usepackage{amsfonts}
\usepackage{amssymb}
\usepackage{eucal}
\def\({\left(}
\def\){\right)}
\newcommand{\bra}[1]{\langle #1 |}        
\newcommand{\ket}[1]{{| #1 \rangle}}      

\newcommand{\nn}{\nonumber}
\newcommand{\bea}{\begin{eqnarray}}
\newcommand{\ena}{\end{eqnarray}}
\newcommand{\be}{\begin{eqnarray*}}
\newcommand{\en}{\end{eqnarray*}}
\newcommand{\ba}{\begin{array}}
\newcommand{\ea}{\end{array}}
\def\og{\overline{\gamma}}
\def\P{\mathcal P}

\newcommand{\R}{{\mathbb R}}
\newcommand{\C}{{\mathbb C}}
\newcommand{\Z}{{\mathbb Z}} 
 
\newcommand{\cP}{\mathcal{P}}

\newcommand{\cK}{\mathcal{K}}
\newcommand{\slt}{\mathfrak{sl}_2}
\newcommand{\res}{{\rm res}}
\newcommand{\id}{{\rm id}}
\newcommand{\tr}{{\rm tr}}
\newcommand{\Tr}{{\rm Tr}}
\newcommand{\End}{\mathop{\rm End}}
\newenvironment{tenumerate}{
  \begin{enumerate}
  
  }{\end{enumerate}}
\newcommand{\bi}{\begin{tenumerate}}
\newcommand{\ei}{\end{tenumerate}}
\newcommand{\isoto}[1][]%
{{\mathop{\buildrel{\sim}\over\longrightarrow}\limits_{#1}}}

\newcommand{\la}{\lambda}
\newcommand{\g}{\gamma}
\newcommand{\al}{\alpha}
\newcommand{\e}{\epsilon}
\newcommand{\eb}{\bar{\epsilon}}

\numberwithin{equation}{section}
\newtheorem{thm}{Theorem}[section]
\newtheorem{prop}[thm]{Proposition}
\newtheorem{lem}[thm]{Lemma}

\newtheorem{cor}[thm]{Corollary}

\begin{document} 
\title[Correlation functions of XXX model]
{A recursion formula for the correlation 
functions of an inhomogeneous XXX model}
\author{H.~Boos, M.~Jimbo, T.~Miwa, F.~Smirnov and Y.~Takeyama}
\address{HB: Physics Department, University of Wuppertal, D-42097,
Wuppertal, Germany\footnote{
on leave of absence from the Institute for High Energy Physics, Protvino, 
142281, Russia}}\email{boos@physik.uni-wuppertal.de}
\address{MJ: Graduate School of Mathematical Sciences, The
University of Tokyo, Tokyo 153-8914, Japan}\email{jimbomic@ms.u-tokyo.ac.jp}
\address{TM: Department of Mathematics, Graduate School of Science,
Kyoto University, Kyoto 606-8502, 
Japan}\email{tetsuji@math.kyoto-u.ac.jp}
\address{FS\footnote{Membre du CNRS}: Laboratoire de Physique Th{\'e}orique et
Hautes Energies, Universit{\'e} Pierre et Marie Curie,
Tour 16 1$^{\rm er}$ {\'e}tage, 4 Place Jussieu
75252 Paris Cedex 05, France}\email{smirnov@lpthe.jussieu.fr}
\address{YT: Department of Mathematics, 
Tsukuba University, Tsukuba 305-8571, Japan}
\email{takeyama@math.tsukuba.ac.jp}

\date{\today}
\dedicatory{Dedicated to Ludwig Faddeev on the occasion of 
his seventieth birthday}
\begin{abstract}
A new recursion formula is presented for the correlation functions of the 
integrable spin $1/2$ XXX chain with inhomogeneity.  It relates the
correlators involving $n$ consecutive lattice sites to those with $n-1$ and
$n-2$ sites. In a series of papers by V.~Korepin and two of the present
authors, it was discovered that the correlators have a certain specific
structure as functions of the inhomogeneity parameters. Our formula allows
for a direct proof of this structure, as well as an exact description of
the rational functions which has been left undetermined in the previous works. 
\end{abstract}
\maketitle
\setcounter{section}{0}
\setcounter{equation}{0}
\tableofcontents
%
%
%
%
%
%
%
%
%
%
%
%
%
%
%
%
%
\section{Introduction}\label{sec:1}
Consider the XXX anti-ferromagnet given by the Hamiltonian
\be
H_{XXX}=\frac{1}{2}\sum_j
\left(\sigma_j^x\sigma_{j+1}^x
+\sigma_j^y\sigma_{j+1}^y+\sigma_j^z\sigma_{j+1}^z\right).
\en
This model was solved in the famous paper by Bethe \cite{Bethe} 
already in 1931, using
what is now called the coordinate Bethe Ansatz. 
Nevertheless it took some time before
the physical content of the model in the thermodynamic limit 
was completely clarified. 
The spectrum of excitations was correctly described 
for the first time in the paper 
by Faddeev and Takhtajan \cite{FT}; 
it was shown that the spectrum contains magnons of spin $1/2$. 
These authors used the algebraic Bethe Ansatz 
formulated by Faddeev, Sklyanin and Takhtajan (see the review \cite{ABA})  
on the basis of $R$-matrices and the Yang-Baxter equation. 
The origin of these new techniques goes back to the works of Baxter \cite{bax}. 

Let us recall on our example
the role of $R$-matrices in solvable models. 
The XXX model is related to the rational $R$-matrix 
which acts on $\mathbb{C}^2\otimes \mathbb{C}^2$ 
(see \eqref{eq:R} for the explicit formula). 
We use the usual notation $R_{1,2}(\la)$ 
where $1,2$ label the corresponding spaces and 
$\la$ is the spectral parameter. 
The relation of the $R$-matrix to the 
XXX Hamiltonian is as follows. 
Consider the transfer matrix 
\be
t_N(\la )
=\tr \(R_{\alpha,-N}(\la)R_{\alpha,-N+1}(\la)\cdots R_{\alpha,N}(\la )\),
\en
where the trace is 
taken with respect to the auxiliary space labelled by $\alpha$.
The transfer matrices commute for different values of 
spectral parameters, 
giving rise to a commuting family of operators. 
If we expand $t_N(\la )$ in powers of $\la$, 
\be
\log t_N(0)^{-1}t_N(\la )=\sum\limits _{k=1}^{\infty} I_k\la ^k,   
\en
then $I_1$ coincides with the XXX Hamiltonian for the 
periodic chain of length $2N+1$. 
All other $I_k$ are integrals of motion commuting with $I_1$. 

We have repeated these well-known facts in order to make clear 
the following remark due to Baxter. 
Consider the inhomogeneous chain whose integrals of motion 
are generated by the transfer matrix 
\be
\tr \(R_{a,-N}(\la)\cdots R_{a,0}(\la)
R_{a,1}(\la -\la _1)\cdots R_{a,n}(\la -\la _n) 
R_{a,n+1}(\la)\cdots R_{a,N}(\la )\), 
\en
where $\la_1,\cdots \la _n$ are arbitrary parameters. 
The model is still exactly solvable, 
although the interaction is not local any more.
This generalization will be very important for us.

After the calculation of the spectrum, the next important 
issue is that of the correlation functions.  
Following \cite{JM}, we consider general correlators of the form
\be
\langle {\rm vac}|
(E_{\e_1,\eb_1})_1\cdots (E_{\e_n,\eb_n})_n
|{\rm vac} \rangle.  
\en
These are the averages over the ground state $|{\rm vac}\rangle$ 
of products of elementary operators $(E_{\e_j,\eb_j})_j$ 
($\e_j,\eb_j=\pm 1$) at site $j$. 
For $n=2$, they can be calculated easily from the vacuum energy. 
The first non-trivial result is due to Takahashi \cite{Tak}, 
who evaluated the correlators for $n=3$ 
in terms of $\zeta(3)$, where $\zeta$ 
is the Riemann $\zeta$-function. 

Results for general $n$ were brought forth 15 years later 
by Jimbo, Miki, Miwa and Nakayashiki \cite{JMMN}, 
in the framework of representation theory of quantum affine algebras. 
Their results and further developments 
are presented in the book \cite{JM}. 
It was shown in the context of a more general XXZ model that 
the correlators in both homogeneous and 
inhomogeneous cases are given in terms 
of multiple integrals in which the number of 
integrations is equal to the distance $n$ on the lattice.  
Actually, the inhomogeneous case plays a very 
important role. The correlators in this case depend 
on the parameters $\la _j$ since the vacuum does. 
As functions of $\la _j$, 
the correlators satisfy the 
quantum Knizhnik-Zamolodchikov equation (qKZ) \cite{FR}
with level $-4$.   
Here an unexpected similarity became apparent  
between the correlators in lattice models and 
form factors in integrable relativistic models 
calculated by Smirnov \cite{Sbk}. 
The latter also satisfy the qKZ equation but with level $0$. 
The symmetry algebra of the XXX model is the Yangian, 
and the qKZ equation in this situation has been 
discussed in \cite{SmY}.

One remark is in order here. 
The algebraic methods of \cite{JM} work nicely in the presence 
of a gap in the spectrum. 
In the gapless case (such as the XXX model under consideration),  
the formulae for the correlators were obtained by 
"analytical continuation". 
However, later the same formulae were derived rigorously 
by Kitanine, Maillet and Terras \cite{KMT} on the basis of 
the algebraic Bethe Ansatz. 

Actually, it is not very simple to 
obtain the result of Takahashi from
the formulas in \cite{JMMN}. 
For $n=3$, we have three-fold integrals, 
and the result must be surprisingly simple. 
This observation was the starting point of the works 
by Boos and Korepin \cite{BK}. 
They were able to calculate further the cases $n=4$ and $n=5$ 
(for some correlators).  
In all cases the multiple integrals mysteriously disappeared, 
and the results were expressed in terms of products of $\zeta$-functions
at odd positive integers with rational coefficients. 
This mystery had to be explained.

The explanation was presented in the series of
papers by Boos, Korepin and Smirnov \cite{BKS1,BKS2,BKS3}.
The idea is again the use of inhomogeneous model and the qKZ equation. 
In the paper \cite{BKS1} was put forward a conjecture which states
that all the correlators are expressed schematically as follows.
\begin{align}
\langle {\rm vac}|
(E_{\e_1,\eb_1})_1\cdots (E_{\e_n,\eb_n})_n
|{\rm vac} \rangle=
\sum \prod\omega(\la _i-\la_j)\, 
f(\la _1,\cdots ,\la _n).
\label{anz}
\end{align}
Here $\omega(\la)$ is a certain transcendental function 
(a linear combination of logarithmic
derivatives of the $\Gamma$-function), the 
$f(\la _1,\cdots ,\la _n)$ are 
rational functions, and the sum is taken over partitions 
(for the precise formula, see \eqref{eq:Ansatz} and
Theorem \ref{prop:ansatz1}, \ref{prop:ansatz2}). 
In the homogeneous limit, odd integer values of the 
$\zeta$-function appear as the coefficients in the Taylor series 
of $\omega(\la)$.

It was explained that the real reason for the formula (\ref{anz}) 
to hold is  the existence of a formula for 
solutions to the qKZ equation with level $-4$ 
different than those given in \cite{JM}.
This follows from a duality between the 
level $0$ and level $-4$ cases.
In the case of level $0$,  
all solutions in the $\slt$-invariant subspace     
are known. 
To get solutions for the level $-4$ case, 
one has to invert the matrix of solutions 
for the level $0$ case.  
This can be done efficiently 
due to the relation to the symplectic group. 
Detailed explanations of this point will 
bring us too far from the subject of the
 present paper.

Let us consider (\ref{anz}) as an Ansatz. 
The question is to calculate the rational functions 
$f(\la _1,\cdots ,\la _n)$. 
It turns out that this problem is much more 
complicated than it appears at a first glance. 
Despite the efforts made in the papers \cite{BKS1,BKS2,BKS3},  
only partial answers were gained for small $n$:
All the correlation functions for the homogeneous case of the XXX model 
until $n=4$ were calculated by K.Sakai, M.Shiroishi, Y.Nishiyama and
M.Takahashi \cite{SSNT} while the result for 
those functions in the XXZ case was done in the 
papers \cite{KSTS1}, \cite{TKS}, \cite{KSTS2}. 
The whole set of correlation functions 
for $n=5$  both in homogeneous and inhomogeneous cases has been
recently found in \cite{BST}.

The main results of the present paper are
\begin{enumerate}
\item Calculation of the functions $f(\la _1,\cdots ,\la _n)$, 
\item Proof of the Ansatz (\ref{anz}).
\end{enumerate}
\noindent
Surprisingly enough, the result is expressed in terms 
of transfer matrices over an auxiliary 
space of `fractional dimension'. 
This brings us very close to the theory of the Baxter 
$Q$-operators developed by Bazhanov, Lukyanov and 
Zamolodchikov \cite{BLZ}. 
We hope to discuss this issue in future publications. 

The text is organized as follows. 

We begin in 2.1 by formulating the problem of computing the
correlation functions. We review the Ansatz of \cite{BKS1,BKS2,BKS3}. 
In 2.2 we prepare basic materials 
{}from the algebraic Bethe Ansatz.
In 2.3 we summarize the properties of the solution to the qKZ equation
relevant to the correlators. 

We introduce in 3.1 the trace over a space of `fractional dimension'. 
Using this notion, we define in 3.2 the rational function $X^{[i,j]}$ which 
enter the recursion formula as the coefficients. 
The recursion is stated in 3.3. Simple examples are given in 3.4 
for the correlation functions obtained from the recursion.

In 4.1, we give several properties of $X^{[i,j]}$. Using these properties and
the recursion, we prove the Ansatz in 4.2. The proof of recursion is started in
4.3 by calculating the residues of the correlation functions.
The proof is completed in 4.4 by giving an asymptotic estimate.

In Appendix A we discuss the relations between the different 
gauges of the Hamiltonian used here and in the literature. 
In Appendix B we give a proof of the analytic and 
asymptotic properties of the correlators
used in the text.

\section{Correlation functions and reduced qKZ equation}\label{sec:2}

\subsection{Formulation of the problem}\label{subsec:2.1}
In this subsection, we formulate the problem we are going to address. 

Consider the XXX model with the Hamiltonian 
\begin{equation}
H_{XXX}=\frac{1}{2}\sum_j
\left(\sigma_j^x\sigma_{j+1}^x
+\sigma_j^y\sigma_{j+1}^y+\sigma_j^z\sigma_{j+1}^z\right).\label{APPA1}
\end{equation}
Its integrability is due to the Yang-Baxter relation. 
The problem is to calculate the correlation functions
\be
\langle {\rm vac}|
(E_{\e_1,\eb_1})_1\cdots (E_{\e_n,\eb_n})_n
|{\rm vac} \rangle.  
\en
These are the averages over the ground state $|{\rm vac}\rangle$ 
of products of elementary operators 
\be
E_{\e_j,\eb_j}=
\Bigl(\delta_{\e_j\alpha}\delta_{\eb_j\beta}\Bigr)_{\alpha,\beta=\pm}
\qquad (\e_j,\eb_j=\pm) 
\en 
acting on the site $j$. 
More precisely, we consider the correlation functions 
of an inhomogeneous model, in which 
each site $j$ carries an independent spectral parameter $\la_j$. 
An exact integral formula for these quantities 
has been known \cite{JM,KMT}. 

It is convenient to pass to the quantity
\begin{equation}
h_n(\la_1,\cdots,\la_n)^{-\e_1,\cdots,-\e_n,\eb_n,\cdots,\eb_1}
=(-1)^{[n/2]}\prod_{j=1}^n(-\eb_j)
\langle{\rm vac}|(E_{\e_1,\eb_1})_1\cdots (E_{\e_n,\eb_n})_n|{\rm vac}\rangle.
\label{APPA2}
\end{equation}
In Appendix A, we give a connection of this formula to a similar formula
for the XXZ model given in \cite{JM}.

Denoting by $v_+,v_-$ the standard basis of
\[
V=\C^2=\C v_+\oplus \C v_-, 
\]
we regard 
\be
&&h_n(\la_1,\cdots,\la_n)
\\
&&\quad
=
\sum_{\e_1,\cdots,\e_n,\eb_1,\cdots,\eb_n}
h_n(\la_1,\cdots,\la_n)^{\e_1,\cdots,\e_n,\eb_n,\cdots,\eb_1}
v_{\e_1}\otimes \cdots v_{\e_n}\otimes v_{\eb_n}\otimes\cdots \otimes
v_{\eb_1}
\en
as an element of $V^{\otimes 2n}$. 
This function is obtained from a 
solution of the qKZ equation with level $-4$
\be
g_{2n}(\lambda_1,\cdots,\lambda_{2n})
\in V^{\otimes 2n}
\en
by specializing the arguments as (see e.g. \cite{JM})
\bea
h_n(\la_1,\cdots,\la_n)=
g_{2n}(\la_1,\cdots,\la_n,\la_n+1,\cdots,\la_1+1).
\label{eq:special}
\ena
In \cite{BKS1,BKS2}, it was found that the functions 
$h_n$ have the following structure.  
\bea
&&h_n(\la_1,\cdots,\la_n)
=\sum_{m=0}^{[n/2]}\sum_{I,J}
\prod_{p=1}^m \omega(\la_{i_p}-\la_{j_p})\, 
f_{n,I,J}(\la_1,\cdots,\la_n).
\label{eq:Ansatz}
\ena
Here $I=(i_1,\ldots,i_m)$, $J=(j_1,\ldots,j_m)$ and
the sum is taken over all sequences $I,J$ such that $I\cap J=\emptyset$,
$1\le i_p<j_p\le n$ $(1\leq p\leq m)$ and $i_1<\cdots<i_m$.

A characteristic feature of 
the formula \eqref{eq:Ansatz} is that the transcendental functions 
enter only through the single function
\footnote{The function $\omega(\la)$ is related to $G(\la)$ in \cite{BKS1}
by $\omega(\la)=G(i\la)+1/2$.}
\bea
&&\omega(\la)=(\la^2-1)\frac{d}{d\la}\log\rho(\la)+\frac{1}{2}
=\sum_{k=1}^\infty(-1)^k\frac{2k(\la^2-1)}{\la^2-k^2}
+\frac{1}{2},
\label{eq:omega}
\ena
where
\bea
&&
\rho(\la)=-\frac{\Gamma\left(\frac{\la}{2}\right)
\Gamma\left(-\frac{\la}{2}+\frac{1}{2}\right)}
{\Gamma\left(-\frac{\la}{2}\right)
\Gamma\left(\frac{\la}{2}+\frac{1}{2}\right)}\,.
\label{eq:rho}
\ena
The remaining factors $f_{n,I,J}(\la_1,\cdots,\la_n)$ 
are rational functions of $\la_1,\cdots,\la_n$ 
with only simple poles along the diagonal $\la_i=\la_j$.
Subsequently it was explained \cite{BKS2,BKS3} 
that this structure of $h_n$ originates from a duality between 
solutions of the qKZ equations with level $0$ and level $-4$. 
On the basis of this relation, \eqref{eq:Ansatz} 
was derived under certain assumptions. 
In these works, the rational functions 
$f_{n,I,J}$ have been left undetermined. 
Our goal in this paper is to derive a recursion formula for $h_n$, 
which enables us to describe $f_{n,I,J}$ and 
to give a direct proof of \eqref{eq:Ansatz}. 

\subsection{$L$ operators and fusion of $R$ matrices}\label{subsec:2.2}
In this subsection, we introduce our notation concerning
$R$ matrices and $L$ operators, and give several formulae for the
fusion of $R$ matrices.

The basic $R$ matrix relevant to the XXX model is 
\begin{equation}
R(\la)=\rho(\la)\frac{r(\la)}{\la+1},\label{eq:R}
\end{equation}
where $\rho(\la)$ is given by \eqref{eq:rho}, 
\be
&&
r(\la)=\la+P,
\en
and $P\in\End\bigl((\C^2)^{\otimes 2}\bigr)$ is 
the permutation operator, $P(u\otimes v)=v\otimes u$. 

We consider also $R$ matrices 
associated with higher dimensional representations of $\slt$. 
Let us fix the convention as follows. 
We denote by 
\be
\pi^{(k)}\,:\,U(\slt)\longrightarrow \End(V^{(k)}), 
\quad V^{(k)}\simeq\C^{k+1}
\en 
the $(k+1)$-dimensional irreducible representation of $\slt$. 
We choose a basis $\{v^{(k)}_j\}_{j=0}^k$ of $V^{(k)}$ 
on which the standard generators $E,F,H$ act as 
\be
E v^{(k)}_j=(k-j+1) v^{(k)}_{j-1},
\quad 
F v^{(k)}_j=(j+1) v^{(k)}_{j+1},
\quad 
H v^{(k)}_j=(k-2j) v^{(k)}_j
\en
with $v^{(k)}_{-1}=v^{(k)}_{k+1}=0$.
{}For $k=1$, we also write $V=V^{(1)}$ and 
$v_+=v^{(1)}_0$, $v_-=v^{(1)}_1$. 
We shall use the singlet vectors in $V^{(k)}\otimes V^{(k)}$ 
($k=1,2$) normalized as 
\bea
&&s^{(1)}=v_+\otimes v_- -v_-\otimes v_+,
\label{eq:singl1}\\
&&s^{(2)}=v^{(2)}_0\otimes v^{(2)}_2-\frac{1}{2}v^{(2)}_1\otimes v^{(2)}_1
+v^{(2)}_2\otimes v^{(2)}_0.
\label{eq:singl2}
\ena

Let $\{S_a\}_{a=1}^3$, $\{S^a\}_{a=1}^3$ be a dual basis of $\slt$ 
with respect to the invariant bilinear form $(x|y)$ normalized as $(H|H)=2$.
It is  well known that the element 
\bea
L^{(1)}(\la)=\la+\frac{1}{2}+\sum_{a=1}^3S_a\otimes \pi^{(1)}(S^a)
\quad \in U(\slt)\otimes\End(V^{(1)})
\label{eq:Lop}
\ena
is a solution of the Yang-Baxter relation
\be
R_{1,2}(\la_1-\la_2)L^{(1)}_1(\la_1)L^{(1)}_2(\la_2)
=
L^{(1)}_2(\la_2)L^{(1)}_1(\la_1)R_{1,2}(\la_1-\la_2).
\en
The suffix stands for the tensor components on which the operators 
act non-trivially. We also use a suffix of the form
$(\alpha_1,\ldots,\alpha_k)$ to denote the symmetric part
$V^{(k)}\subset V_{\alpha_1}\otimes\cdots\otimes V_{\alpha_k}$,
where $V_{\alpha_i}$ are copies of $V$. 

Identifying $V^{(2)}$ with the subspace 
$V^{(2)}_{(1,2)}=\cP^+_{1,2}\left(V_1 \otimes V_2 \right)$ 
where $\cP^{\pm}=(1/2)(1\pm P)$, we define the fused $L$-operator
\bea
L^{(2)}_{(1,2)}(\la)=
L^{(1)}_1\Bigl(\la-\frac{1}{2}\Bigr)
L^{(1)}_2\Bigl(\la+\frac{1}{2}\Bigr)\cP^+_{1,2}
\quad \in U(\slt)\otimes\End(V^{(2)}). 
\label{eq:Lop2}
\ena
Explicitly it is given by
\begin{eqnarray}\label{L^2}
&&L^{(2)}(\la)=\la(\la+1)-\frac{1}{2}C\otimes \id_{V^{(2)}}\\
&&+(\la+1)\sum_{a=1}^3S_a\otimes \pi^{(2)}(S^a)
+\frac{1}{2}\sum_{a,b=1}^3 S_aS_b\otimes\pi^{(2)}\bigl(S^aS^b\bigr).\nn
\end{eqnarray}
In the right hand side, 
\begin{equation}
C=\sum_{a=1}^3 S_aS^a
\end{equation}
denotes the Casimir operator. 

We shall make use of the crossing symmetry
\begin{equation}\label{eq:crossing}
L^{(k)}_\alpha(\la)s^{(k)}_{\alpha,\beta}=
L^{(k)}_\beta(-\la-1)s^{(k)}_{\beta,\alpha}
\end{equation}
and the quantum determinant relation 
\begin{equation}\label{SINGL}
\P^-_{1,2}L^{(1)}_1(\lambda-1)L^{(1)}_2(\lambda)
=\left(\lambda^2-\frac14-\frac12C\right)\P^-_{1,2}.
\end{equation}

Taking the images of \eqref{eq:Lop}, \eqref{eq:Lop2}  
in $V^{(k)}$, we obtain (numerical) $R$ matrices
\footnote{We have $r^{(1,2)}_{3,(12)}(\lambda)=(\lambda+\frac12)
r^{(2,1)}_{(12),3}(\lambda)$.}
\begin{equation}\label{RKL}
r^{(k,l)}(\la)=(\pi^{(k)}\otimes\id)L^{(l)}(\la)
\quad \in \End\bigl(V^{(k)}\otimes V^{(l)}\bigr).
\end{equation}
In the following, we abbreviate $L^{(1)}(\lambda)$ to $L(\lambda)$.

We prepare several formulas about the 
fusion of $R$ matrices.
We have
\begin{equation}\label{L4}
r_{1,\alpha}(\lambda-\frac{k-1}2)
r_{2,\alpha}(\lambda-\frac{k-3}2)
\cdots
r_{k,\alpha}(\lambda+\frac{k-1}2)
\P^+_{1,\cdots, k}=
c_k(\lambda)r^{(k,1)}_{(1,\cdots, k),\alpha}(\lambda)
\end{equation}
where
\begin{equation}
c_k(\lambda)=\prod_{j=1}^{k-1}(\lambda-\frac{k-1}2+j).
\end{equation}
We have also for all $k\ge 1$
\begin{equation}\label{L6}
r^{(k,1)}_{(\al_1,\cdots,\al_k),\alpha}(\lambda-\frac12)
r^{(k,1)}_{(\al_1,\cdots,\al_k),\beta}(\lambda+\frac12)
\P^+_{\alpha\beta}=r^{(k,2)}_{(\al_1,\cdots,\al_k),(\alpha\beta)}
(\lambda)
\end{equation}
and the relation
\begin{eqnarray}\label{L9}
&&(\lambda+\frac{k_1-k_2}2)(\lambda+1+\frac{k_1-k_2}2)
r^{(k_1+k_2,2)}_{(1,\ldots,k_1+k_2),(\alpha\beta)}(\lambda)\\
&&=r^{(k_1,2)}_{(1,\ldots,k_1),(\alpha\beta)}(\lambda-\frac{k_2}2)
r^{(k_2,2)}_{(k_1+1,\ldots,k_1+k_2),(\alpha\beta)}(\lambda+\frac{k_1}2).\nn
\end{eqnarray}


\subsection{Properties of $h_n(\lambda_1,\ldots,\lambda_n)$}\label{subsec:3.1}
The function $h_{n}$ is given by 
a specialization \eqref{eq:special} of the solution $g_{2n}$ 
of the qKZ equation with level $-4$. 
In this subsection, 
we summarize the properties of $h_{n}$ implied by the ones of $g_{2n}$.

In what follows, we deal with various vectors in the tensor product 
$V^{\otimes 2n}$ and those obtained by permuting the tensor components.
In order to simplify the presentation, we adopt the following convention. 
Consider the tensor product 
\begin{equation}
W=V_1\otimes\cdots\otimes V_n\otimes 
V_{\bar{n}}\otimes \cdots\otimes V_{\bar{1}}
\label{SPACE}
\end{equation}
of $2n$ copies of $V$ labeled by $1,\ldots,n,\bar{n},\ldots,\bar{1}$.
For a vector
\begin{equation}
f=\sum f^{\epsilon_1,\ldots,\epsilon_n,\bar{\epsilon}_n,
\ldots,\bar{\epsilon}_1}
v_{\epsilon_1}\otimes\ldots\otimes v_{\epsilon_n}\otimes
v_{\bar{\epsilon}_n}\otimes\ldots\otimes v_{\bar{\epsilon}_1}
\label{VECTORF}
\end{equation}
in $V^{\otimes2n}$ we denote by $f_{1,\ldots,n,\bar n,\ldots,\bar1}$
the same vector (\ref{VECTORF}) in (\ref{SPACE}).
Vectors obtained by permuting tensor components will be indicated 
by permuting suffixes from the `standard position'. 
For example, if 
\be
f=\sum f^{\e_1,\e_2,\eb_2,\eb_1}
v_{\e_1}\otimes v_{\e_2}\otimes v_{\eb_2}\otimes v_{\eb_1}\in V^{\otimes4},
\en
then $f_{2,\bar{2},1,\bar{1}}\in
V_1\otimes V_2\otimes V_{\bar2}\otimes V_{\bar1}$ is given by
\be
f_{2,\bar{2},1,\bar{1}}
&=&P_{1,\bar{2}}P_{1,2}f_{1,2,\bar{2},\bar{1}}
\\
&=&\sum f^{\e_2,\eb_2,\e_1,\eb_1}
v_{\e_1}\otimes v_{\e_2}\otimes v_{\eb_2}\otimes v_{\eb_1}.
\en
By $f_{2,\bar{2},1,\bar{1}}$ we do not mean a vector
in $V_2\otimes V_{\bar2}\otimes V_1\otimes V_{\bar1}$.
Notice that in this notation 
$s^{(1)}_{\bar{1},1}=-s^{(1)}_{1,\bar{1}}\in V_1\otimes V_{\bar1}$. 
Sometimes, we need to construct a vector in 
$V_1\otimes V_2\otimes V_{\bar2}\otimes V_{\bar1}$
{}from one in $V_1\otimes V_{\bar1}$ and one in $V_2\otimes V_{\bar2}$.
Suppose that $f,g\in V\otimes V$. Then, we have
$f_{1,\bar1}\in V_1\otimes V_{\bar1}$, and
$g_{2,\bar2}\in V_2\otimes V_{\bar2}$. We write
$f_{1,\bar1}g_{2,\bar2}$ for the vector
\[
\sum f^{\e_1,\eb_1}g^{\e_2,\eb_2}v_{\e_1}\otimes
v_{\e_2}\otimes v_{\eb_2}\otimes v_{\eb_1}\in
V_1\otimes V_2\otimes V_{\bar2}\otimes V_{\bar1}.
\]
In this convention, ordering of the tensor product is irrelevant.
There is no preference in writing $f_{1,\bar1}g_{2,\bar2}$
or $g_{2,\bar2}f_{1,\bar1}$ to represent the above vector.

We have three more remarks. First,
we use ``auxiliary'' spaces in addition to the ``quantum'' spaces
$V_1,V_{\bar1},\ldots,V_n,V_{\bar n}$. We use $\alpha,\beta,\alpha_1,\alpha_2$,
etc., to label these spaces.
Second, we use the index with parenthesis like
$(\alpha_1,\ldots,\alpha_k)$ to label
the completely symmetric subspace $V^{(k)}\subset V^{\otimes k}$.
This was already mentioned in Section \ref{subsec:2.2}.
If $f$ is a vector in this subspace, we denote by 
$f_{(\alpha_1,\ldots,\alpha_k)}$
the corresponding vector in $V_{\alpha_1}\otimes\cdots\otimes V_{\alpha_k}$.
Lastly, we use similar convention for matrices as well.

The function $g_{2n}$ has the following properties \cite{JM}: 
\bea
&&
g_{2n}(\la_1,\cdots,\la_{2n}) 
\mbox{ is invariant under the action of $\slt$ },
\\
&&
g_{2n}(\cdots,\la_{j+1},\la_j,\cdots)_{\cdots , j+1, j, \cdots}
=R_{j,j+1}(\la_{j,j+1})
g_{2n}(\cdots , \la_{j}, \la_{j+1}, \cdots)_{\cdots , j, j+1, \cdots}, 
\label{eq:g-qKZ1} \\
&&
g_{2n}(\la_{1}, \cdots , \la_{2n-1}, \la_{2n}+2)_{1, \cdots , 2n}=
(-1)^{n}g_{2n}(\la_{2n}, \la_{1}, \cdots , \la_{2n-1})_{2n, 1, \cdots , 2n-1}, 
\\ 
&& 
g_{2n}(\la_{1}, \cdots , \la_{2n-2}, \la, \la-1)_{1, \cdots , 2n}=
g_{2n-2}(\la_{1}, \cdots, \la_{2n-2})_{1, \cdots , 2n-2}\,
s_{2n-1, 2n}^{(1)}.
\label{eq:g-qKZ2} 
\ena 
Note that $R(-1)=-1+P=-2\cP^-$ and $s^{(1)}_{21}=-s^{(1)}_{12}$.
{}From this, \eqref{eq:g-qKZ1} and \eqref{eq:g-qKZ2}, we have 
\be 
\mathcal{P}_{2n-1,2n}^{-}
g_{2n}(\la_{1}, \cdots , \la_{2n-2}, \la, \la+1)_{1, \cdots , 2n}
=-\frac{1}{2}
g_{2n-2}(\la_{1}, \cdots , \la_{2n-2})_{1, \cdots , 2n-2} 
\, s_{2n-1, 2n}^{(1)}. 
\en

Here and after we set $\lambda_{i,j}=\lambda_i-\lambda_j$. 
For $\alpha=1$ or $\bar1$, set
\bea
&&A_\alpha(\la_1,\cdots,\la_n)\label{FUNA}\\
&&=(-1)^n
R_{\alpha,\bar{2}}(\lambda_{1,2}-1)\cdots R_{\alpha,\bar{n}}(\lambda_{1,n}-1)
R_{\alpha,n}(\lambda_{1,n})\cdots R_{\alpha,2}(\lambda_{1,2}).\nn
\ena

It is easy to prove the following proposition from the formulae above: 
\begin{prop} 
The function $h_n(\la_1,\cdots,\la_n)$ satisfies the following properties:
\bea
&&
h_n(\la_1,\cdots,\la_n) 
\mbox{ is invariant under the action of $\slt$ },
\label{eq:sl2_inv}
\ena
\bea
&&h_n(\cdots,\la_{j+1},\la_j,\cdots)_{\cdots  , j+1, j, \cdots , \bar{j}, \overline{j+1}, \cdots}
\label{eq:Rsymm}\\
&&{}=
R_{j,j+1}(\la_{j,j+1})
R_{\overline{j+1},\bar{j}}(\la_{j+1,j})
h_n(\cdots,\la_j,\la_{j+1},\cdots)_{\cdots , j, j+1, \cdots , \overline{j+1}, \bar{j}, \cdots},\nn
\ena
\bea
&&
h_n(\la_1-1,\la_2,\cdots,\la_n)_{1, \cdots ,n, \bar{n}, \cdots , \bar{1}}
\label{eq:rqKZ}\\ 
&&{}=A_{\bar1}(\la_1,\cdots,\la_n)
h_n(\la_1,\la_2,\cdots,\la_n)_{\bar{1},2,\cdots,n,\bar{n},\cdots,\bar{2},1},\nn
\ena
\bea
&&
\cP^-_{1,2}\cdot h_n(\la-1,\la,\cdots,\la_n)_{1, \cdots , n, \bar{n}, \cdots \bar{1}} 
{}=-\frac{1}{2}s^{(1)}_{1,2}\,s^{(1)}_{\bar{1},\bar{2}} \, 
h_{n-2}(\la_3,\cdots,\la_n)_{3, \cdots , n, \bar{n}, \cdots , \bar{3}},
\label{eq:n_to_n-2}\\
&&
\cP^-_{1,\bar{1}}\cdot h_n(\la_{1},\cdots,\la_n)_{1, \cdots , n, \bar{n}, \cdots , \bar{1}}
=(-1)^{n-1}\frac{1}{2}s^{(1)}_{1,\bar{1}} \,
h_{n-1}(\la_2,\cdots,\la_n)_{2, \cdots , n, \bar{n}, \cdots , \bar{2}}.
\label{eq:n_to_n-1}
\ena
\end{prop} 
In particular, \eqref{eq:rqKZ} is a reduced form of the qKZ equation.
Notice that, in contrast to the equation \eqref{eq:g-qKZ1} for $g_{2n}$,  
the coefficients appearing in \eqref{eq:Rsymm}and \eqref{eq:rqKZ} are 
rational functions.
This is a consequence of the properties $\rho(\la)\rho(-\la)=1$, 
$\rho(\la-1)\rho(\la)=-\la/(\la-1)$. 

An integral formula for the function $h_{n}$ has been 
constructed in \cite{JM,Kor}. 
{}Using that formula, 
we derive in Appendix \ref{app:B} the following analytic properties of $h_{n}$.
\begin{prop} \label{prop:3.2}
The function $h_{n}(\la_{1}, \cdots , \la_{n})$ satisfies the following: 
\bea
&&h_n(\la_1,\cdots,\la_n)
\mbox{ is meromorphic in $\la_1,\cdots,\la_n$ }\label{eq:meromorphy}\\
&&\mbox{ with at most simple
poles at $\la_i-\la_j\in\Z\backslash\{0,\pm 1\}$},\nn\\
&&\mbox{ for any $0<\delta<\pi$ we have }\label{eq:reg_at_infty}
\\
&&\lim_{\la_1\to\infty \atop \la_1\in S_\delta}
h_n(\la_1,\cdots,\la_n)=(-1)^{n-1}\frac{1}{2}s^{(1)}_{1,\bar{1}}
h_{n-1}(\la_2,\cdots,\la_n),\nn\\
&&\mbox{where
$S_\delta=\{\la\in\C\mid\delta<|{\rm arg}\,\lambda|<\pi-\delta\}$}.\nn
\ena
\end{prop} 
As we shall show in the following sections, 
the properties \eqref{eq:sl2_inv}--\eqref{eq:reg_at_infty} 
uniquely determine the functions $h_n$. 


\section{Recursion formula}
In this section we state the main result of this paper:
a recursion formula for the correlation functions. The main
ingredient of the recursion is a transfer matrix with an auxiliary space
of fractional dimension.

\subsection{Trace function}\label{subsec:2.5}
We define ``trace over a space of fractional dimension''. 
By this we mean the unique $\C[x]$ linear map 
\be
\Tr_x~:~U(\slt)\otimes\C[x]\longrightarrow \C[x]
\en
such that for any non-negative integer $k$ we have
\bea
\Tr_{k+1}(A)=\tr_{V^{(k)}}\pi^{(k)}(A)\qquad (A\in U(\slt)).
\label{eq:tr_x}
\ena
Here $\tr$ in the right hand side stands for the usual trace. 

We list some properties of the trace function $\Tr_x$. 
\bea
&&\Tr_x(AB)=\Tr_x(BA),\quad \Tr_x(1)=x,
\label{eq:tr1}\\
&&\Tr_x(A)=0\text{ if $A$ has non-zero weight},\\
&&\Tr_x(e^{z H})=\frac{\sinh (x z)}{\sinh z},
\label{eq:trH}
\\
&&
\Tr_x\left(\Bigl(\frac{H^2}{2}+H+2 FE\Bigr) A\right)
=\frac{x^2-1}2\Tr_x(A)
\qquad (A\in U(\slt)\otimes \C[x]).
\label{eq:trC}
\ena

By the generating series \eqref{eq:trH} the traces $\Tr_x(H^a)$ are known,
and using \eqref{eq:trC} one can calculate $\Tr_x(H^aE^bF^c)$ inductively 
for all $a,b,c\ge 0$. 
We emphasize that $\Tr_x(A)$ is determined by the `dimension' 
$\Tr_x(1)=x$ and the value of the Casimir operator; we have
\bea
\Tr_x(A)=\Tr_x(A')\text{ if $\varpi_x(A)=\varpi_x(A')$}, 
\label{eq:Tr-varpi}
\ena
where $\varpi_x$ is the projection 
\begin{equation}
\varpi_x:U(\frak{sl}_2)\otimes\C[x]
\rightarrow U(\frak{sl}_2)\otimes\C[x]/I_x
\label{eq:varpi}
\end{equation}
and $I_x$ signifies the two-sided ideal of 
$U(\frak{sl}_2)\otimes\C[x]$ generated by $C-(x^2-1)/2$. 

The following are simple consequences of these rules. 
\bea
&&
\Tr_{-x}(A)=-\Tr_x(A),
\label{eq:tr2}\\
&&
\Tr_x(A)-x\varepsilon(A) \in x(x^2-1)\C[x] 
\label{eq:tr3}
\\
&&\qquad 
\mbox{ where $\varepsilon:U(\slt)\otimes\C[x]\to\C[x]$ stands for the counit}, 
\nn\\
&&\mbox{The degree of $\Tr_x(H^aE^bF^c)$ is at most $m+1$ ($m$ even)}
\label{eq:deg_of_Tr}\\
&&\qquad\quad\mbox{ or $m$ ($m$ odd) where $m=a+b+c$}.
\nn
\ena
\subsection{Functions $X^{[i,j]}$}\label{subsec:2.3}
We are now going to introduce our main object $X^{[i,j]}$. 

For $1\le j\le n$ (resp. $1\le i<j\le n$), we set 
\bea
&&
W^{[j]}=V_1\otimes\cdots\overset{j}{\widehat{\phantom{L}}}\cdots\otimes
V_n\otimes V_{\bar{n}}\otimes\cdots\overset{\bar j}{\widehat{\phantom{L}}}\cdots
\otimes V_{\bar{1}}\,,
\label{eq:Wj}
\\
&&
W^{[i,j]}=
V_1\otimes
\cdots\overset{i}{\widehat{\phantom{L}}}\cdots
\cdots\overset{j}{\widehat{\phantom{L}}}\cdots
\otimes 
V_n\otimes 
V_{\bar{n}}\otimes
\cdots\overset{\bar j}{\widehat{\phantom{L}}}\cdots
\cdots\overset{\bar i}{\widehat{\phantom{L}}}\cdots
\otimes V_{\bar{1}}\,.
\label{eq:Wij}
\ena
Define the monodromy matrices
\bea
T^{[j]}(\lambda)
&=&
L_{\bar1}(\la-\la_1-1)\cdots \overset{\bar j}{\widehat{\phantom{L}}}
\cdots L_{\bar n}(\la-\la_n-1)
\label{eq:Tj}
\\
&\times& L_n(\la-\la_n)\cdots \overset{j}{\widehat{\phantom{L}}}
\cdots L_1(\la-\la_1)
\nn\\
T^{[i,j]}(\lambda)
&=&
L_{\bar1}(\la-\la_1-1)
\cdots \overset{\bar i}{\widehat{\phantom{L}}}
\cdots \overset{\bar j}{\widehat{\phantom{L}}}
\cdots L_{\bar n}(\la-\la_n-1)
\label{eq:Tij}\\
&\times&
L_n(\la-\la_n)
\cdots \overset{j}{\widehat{\phantom{L}}}
\cdots \overset{i}{\widehat{\phantom{L}}}\cdots L_1(\la-\la_1).
\nn
\ena
These are elements of
$U(\frak{sl}_2)\otimes\End(W^{[j]})$
and
$U(\frak{sl}_2)\otimes\End(W^{[i,j]})$, 
respectively. 

Using the trace function $\Tr_x$, we define the functions
\be
&&X^{[i,j]}(\la_1,\cdots,\la_n)\in
V_{i}\otimes V_{\bar{i}}\otimes V_j\otimes V_{\bar j}\otimes
\End(W^{[i,j]})
\qquad (1\le i<j\le n)
\en
by the formula 
\bea
\label{eq:Xij}\\
&&X^{[i,j]}(\la_1,\cdots,\la_n)=\frac1{\lambda_{i,j}(\lambda_{i,j}^2-1)
\prod_{p\not=i,j}\lambda_{i,p}\lambda_{j,p}}\nn\\
&&\times R_{i,i-1}(\lambda_{i,i-1})\cdots R_{i,1}(\lambda_{i,1})
R_{\overline{i-1},\bar i}(\lambda_{i-1,i})\cdots
R_{\bar 1\bar i}(\lambda_{1,i})\,
\Tr_{\la_{i,j}}\Bigr(T^{[i]}\bigl(\frac{\lambda_i+\lambda_j}{2}\bigr)\Bigl)\nn\\
&&\times R_{j,j-1}(\lambda_{j,j-1})\cdots 
\widehat{R_{j,i}(\lambda_{j,i})}\cdots R_{j,1}(\lambda_{j,1})
R_{\overline{j-1},\bar{j}}(\lambda_{j-1,j})\cdots 
\widehat{R_{\bar{i},\bar{j}}(\lambda_{i,j})}
\cdots R_{\bar{1},\bar{j}}(\lambda_{1,j})\nn\\
&&\times s^{(2)}_{(i,\bar{i}),(j,\bar{j})}\,.\nn
\ena
One can think of
$\Tr_{\la_{i,j}}(T^{[i]}\bigl(\frac{\lambda_i+\lambda_j}{2}\bigr))$ 
as a transfer matrix with ``$\la_{i,j}$-dimensional auxiliary space''.

The functions $X^{[i,j]}$ are not independent. 
{}From \eqref{eq:Xij} and \eqref{TRANS}, $X^{[i,j]}$ with general $i,j$
can be expressed in terms of one of them, e.g., 
\bea
&&
X^{[1,2]}(\la_1,\cdots,\la_n)=
\frac1{\lambda_{1,2}(\lambda_{1,2}^2-1)
\prod_{p=3}^n\lambda_{1,p}\lambda_{2,p}}
\label{eq:X12}\\
&&\quad \times
\Tr_{\lambda_{1,2}}
\Bigr(T^{[1]}\bigl(\frac{\lambda_1+\lambda_2}{2}\bigr)\Bigl)
s^{(2)}_{(1,\bar{1}),(2,\bar{2})}\,,
\nn
\ena
as follows.
\bea
&&X^{[i,j]}(\la_1,\cdots,\la_n)
\label{eq:XijX12}\\
&=&
R_{i,i-1}(\lambda_{i,i-1})\cdots R_{i,1}(\lambda_{i,1})
R_{j,j-1}(\lambda_{j,j-1})\cdots 
\widehat{R_{j,i}(\lambda_{j,i})}\cdots R_{j,1}(\lambda_{j,1})
\nn
\\
&\times&
R_{\overline{i-1},\bar i}(\lambda_{i-1,i})\cdots
R_{\bar 1\bar i}(\lambda_{1,i})
R_{\overline{j-1},\bar{j}}(\lambda_{j-1,j})\cdots 
\widehat{R_{\bar{i},\bar{j}}(\lambda_{i,j})}
\cdots R_{\bar{1},\bar{j}}(\lambda_{1,j})
\nn
\\
&\times&
X^{[1,2]}(\la_i,\la_j,\la_1,
\cdots,\widehat{\la_i},\cdots,\widehat{\la_j},\cdots,\la_n)
_{i,j,1,\cdots,\hat{i},\cdots,\hat{j},\cdots,n,
\bar{n},\cdots,\hat{\bar{j}},\cdots,\hat{\bar{i}},\cdots,\bar{1},
\bar{j},\bar{i}}.
\nn
\ena
Nonetheless, for the description of the results,   
it is convenient to use all of $X^{[i,j]}$.

Let us list the main properties of $X^{[i,j]}$.
\begin{description}
\item[Transformation law] 
For an element $\sigma$ of the symmetric group $\mathfrak{S}_n$, 
let us use the abbreviation
\be
&&
(X^{[i,j]})^{\sigma}=X^{[i,j]}(\la_{\sigma(1)},\cdots,\la_{\sigma(n)})
_{\sigma(1),\cdots,\sigma(n),\overline{\sigma(n)},\cdots,\overline{\sigma(1)}}
\,.
\en
Then the functions $X^{[i,j]}=(X^{[i,j]})^{\rm id}$ obey
the transformation law
\bea\label{TRANS}
&&
R_{k,k+1}(\la_{k,k+1})R_{\overline{k+1},\bar{k}}(\la_{k+1,k})
X^{[i,j]}
\\
&&\quad
=
\begin{cases}
(X^{[i,j]})^{(k,k+1)}R_{k,k+1}(\la_{k,k+1})R_{\overline{k+1},\bar{k}}(\la_{k+1,k})
& (k\neq i,i-1,j,j-1), \\
(X^{[i,j+1]})^{(j,j+1)}& (k=j), \\
(X^{[i,j-1]})^{(j-1,j)}& (i<k=j-1), \\
(X^{[i-1,j]})^{(i-1,i)}& (k=i-1), \\
(X^{[i+1,j]})^{(i,i+1)}& (k=i<j-1), \\
(X^{[i,i+1]})^{(i,i+1)}& (i=k, j=i+1). \\
\end{cases}
\nn
\ena
\item[Pole structure]
$\prod_{p(\neq i,j)}\la_{i,p}\la_{j,p}\cdot X^{[i,j]}(\la_1,\cdots,\la_n)$ 
is a polynomial. 
\item[Regularity at $\infty$]
For each $1\le k\le n$ and an $\slt$-invariant vector $v\in (W^{[i,j]})^{\slt}$, 
we have
\bea
X^{[i,j]}(\la_1,\cdots,\la_n)v=O(1)
\qquad (\la_k\to\infty).
\label{eq:reg_infty}
\ena
\item[Difference equation]
$X^{[1,2]}(\la_1,\cdots,\la_n)$ satisfies the difference equation
\bea
&&X^{[1,2]}(\la_1-1,\la_2,\cdots,\la_n)
\label{eq:diff_eq}\\
&&
=-\frac{(\lambda_{12}-1)(\lambda_{12}+1)}{\lambda_{12}(\lambda_{12}-2)}
A_{\bar{1}}(\la_1,\cdots,\la_n)P_{1,\bar{1}}
X^{[1,2]}(\la_1,\la_2,\cdots,\la_n)\,,
\nn
\ena
where $A_{\bar{1}}$ is defined in \eqref{FUNA}.
\end{description}
These properties will be proved in Section \ref{sec:4} 
(see Lemma \ref{lem:X2sym}--\ref{lem:degree_of_X} and Lemma \ref{lem:diff_eq}). 

\subsection{Main result}\label{subsec:2.4}
We are now in a position to state the
recursion formula which determines 
the functions $h_n(\la_1,\cdots,\la_n)$
starting from the initial condition
\bea
h_0=1,\quad h_1(\la_1)=\frac{1}{2}s^{(1)}_{1,\bar{1}}\,.
\label{eq:initial}
\ena
The proofs for the statements in this subsection are given in Section 4.

In this subsection we write $X^{[i,j]}_n$ for $X^{[i,j]}$ to
indicate the relevant number of sites $n$. 


\begin{thm}\label{thm:recursion}
We have the following recursion formula. 
\bea
&&h_n(\la_1,\cdots,\la_n)_{1,\cdots,n,\bar{n},\cdots,\bar{1}}
\label{eq:Recursion}
\\
&=& \sum_{j=2}^n Z_n^{[1,j]}(\lambda_1,\cdots,\lambda_n)
\times
h_{n-2}(\la_2,\cdots,\widehat{\la_j},\cdots,\la_n)_
{2,\cdots,\hat{j},\cdots,n,\bar{n},\cdots,\hat{\bar{j}},\cdots,\bar{2}}
\nn\\
&+&
(-1)^{n-1}
\frac{1}{2}s^{(1)}_{1,\bar{1}}\cdot
h_{n-1}(\la_2,\cdots,\la_n)_{2,\cdots,n,\bar{n},\cdots,\bar{2}}.
\nn
\ena
Here 
\bea
Z_n^{[1,j]}(\lambda_1,\cdots,\lambda_n)
&=&\oint_{\mathcal{C}}
\frac{d\sigma}{2\pi i}
\frac{\omega(\sigma-\la_j)}{\sigma-\la_1}
X_n^{[1,j]}(\sigma,\la_2,\cdots,\la_n)\label{eq:Zj}\\
&=&
\omega(\la_{1,j})X_n^{[1,j]}(\la_1,\la_2,\cdots,\la_n)
\nn\\
&+&\sum_{p(\neq 1,j)}\frac{\omega(\la_{p,j})}{\la_{p,1}}
\res_{\sigma=\la_p}X_n^{[1,j]}(\sigma,\la_2,\cdots,\la_n),
\nn
\ena
where $\mathcal{C}$ is a simple closed curve encircling
$\la_1,\cdots,\la_n$ anticlockwise, 
$\omega(\la)$ is given by \eqref{eq:omega}, and
$X_n^{[1,j]}(\la_1,\cdots,\la_n)$ is defined in \eqref{eq:Xij}. 
\end{thm}

\medskip
In the last line of \eqref{eq:Zj}, 
we have used the fact that 
$X_n^{[1,j]}(\sigma,\la_2,\cdots,\la_n)$ has no poles at $\sigma=\la_j$. 

\begin{thm}\label{prop:ansatz1}
The function $h_n(\la_1,\cdots,\la_n)$ has the structure
\bea
&&h_n(\la_1,\cdots,\la_n)
=\sum_{m=0}^{[n/2]}\sum_{I,J} 
\prod_{p=1}^m\omega(\la_{i_p}-\la_{j_p})\,
f_{n,I,J}(\la_1,\cdots,\la_n),
\label{eq:ansatz}
\ena
where $f_{n,I,J}(\la_1,\cdots,\la_n)\in V^{\otimes 2n}$ 
are rational functions, 
and $I=(i_1,\ldots,i_m)$, $J=(j_1,\ldots,j_m)$ 
run over sequences satisfying $I\cap J=\emptyset$, 
$i_1<\cdots<i_m$, $1\le i_p<j_p\le n$ $(1\leq p\leq m)$. 
The representation of $h_n$ in the above form is unique.
\end{thm}

\begin{thm}\label{prop:ansatz2}
In the notation of Theorem \ref{prop:ansatz1}, 
the rational functions $f_{n,I,J}(\la_1,\cdots,\la_n)$ are 
uniquely determined by the recursion relation
\bea
&&
f_{n,iI',jJ'}(\la_1,\cdots,\la_n)=X_n^{[i,j]}(\la_1,\cdots,\la_n)
\label{eq:f5}
\\
&&\quad\times 
f_{n-2,I',J'}
(\la_1,\cdots,\widehat{\la_i},\cdots,\widehat{\la_j},\cdots,\la_n),
\nn
\\
&&
f_{n,\emptyset,\emptyset}(\la_1,\cdots,\la_n)
=(-1)^{n(n-1)/2}\frac{1}{2^n}
s^{(1)}_{1,\bar{1}}\cdots s^{(1)}_{n,\bar{n}}\,.
\label{eq:f6}
\ena
In addition, they enjoy the following properties.   
\bea
&&\quad\mbox{$f_{n,I,J}(\la_1,\cdots,\la_n)$ is invariant 
under the action of $\slt$},\label{eq:f1}\\
&&\quad f_{n,I,J}(\lambda_1,\cdots,\lambda_{j+1},\lambda_j,\cdots,
\lambda_n)_{\cdots,j+1,j,\cdots,\cdots,\overline{j},\overline{j+1},\cdots}
\label{eq:f4}\\
&&=R_{j,j+1}(\lambda_{j,j+1})R_{\overline{j+1},\overline{j}}
(\lambda_{j+1,j})f_{n,\widetilde{I},\widetilde{J}}
(\lambda_1,\cdots,\lambda_j,\lambda_{j+1},\cdots,\lambda_n)
_{\cdots,j,j+1,\cdots,\overline{j+1},\overline{j},\cdots},\nonumber\\
&&\mbox{where $\widetilde{I}=\sigma(I)$ and $\widetilde{J}=\sigma(J)$
$(\sigma=(j,j+1))$}\mbox{ except for the following cases:}\nonumber\\
&&\hbox{if }i_m=j\in I,j_m=j+1\in J\hbox{ for some $m$,
we have }\tilde I=I,\tilde J=J;
\nonumber\\
&&\hbox{if }i_m=j,i_{m+1}=j+1\hbox{ for some $m$, we have }
\tilde I=I\hbox{ and $\tilde J$ is given by }\nonumber\\
&&\tilde j_l=
\begin{cases}
j_l&\hbox{ if }l\not=m,m+1;\\
j_{m+1}&\hbox{ if }l=m;\\
j_m&\hbox{ if }l=m+1,
\end{cases}\nonumber\\
&&\quad\mbox{$f_{n,I,J}(\la_1,\cdots,\la_n)$ is regular
at $\infty$ in each $\la_j$}, 
\label{eq:f2}
\\
&&\quad\mbox{$f_{n,I,J}(\la_1,\cdots,\la_n)$ has at most simple 
poles at $\lambda_i-\lambda_j=0$, where}
\label{eq:f3}
\\
&&1\leq i<j\leq n, ~~\mbox{$i\in I$ or $j\in J$}, ~~
(i,j)\neq (i_p,j_p) (1\leq p\leq m).
\nn
\ena
\end{thm}


In general, $f_{n,I,J}$ is expressed as follows. 
Suppose $I=(i_1,\cdots,i_m)$, $J=(j_1,\cdots,j_m)$
and $\{1,\cdots,n\}\backslash I\cup J=\{k_1,\cdots,k_l\}$, 
where $k_1<\cdots<k_l$, $n=2m+l$. 
Denote the corresponding permutation by
\be
\sigma=
\begin{pmatrix}
1  &2  &\cdots&\cdots  &\cdots&\cdots&\cdots&n  \\
i_1&j_1&\cdots&i_m     &j_m   &k_1   &\cdots&k_l\\
\end{pmatrix}
\quad \in\mathfrak{S}_n, 
\en
and let $\sigma=\sigma_{a_1}\circ\cdots\circ\sigma_{a_N}$ 
be a reduced decomposition into transpositions 
$\sigma_a=(a,a+1)$. 
We associate with $\sigma$ the $R$ matrices
\be
&&
R^\sigma=R_{b'_1,b_1}(\la_{b'_1,b_1})\cdots
R_{b'_N,b_N}(\la_{b'_N,b_N}),
\\
&&
\overline{R}^\sigma=
R_{\overline{b_1},\overline{b'_1}}(\la_{b_1,b'_1})\cdots
R_{\overline{b_N},\overline{b'_N}}(\la_{b_N,b'_N}),
\en
where $b_i=\sigma_{a_1}\cdots\sigma_{a_{i-1}}(a_i)$, 
$b'_i=\sigma_{a_1}\cdots\sigma_{a_{i-1}}(a_i+1)$.  
Then 
\be
&&f_{n,I,J}(\la_1,\cdots,\la_n)
=\frac{1}{D}
R^{\sigma}\overline{R}^{\sigma}
\tau_1\tau_2\cdots \tau_m {\bf s}, 
\en
where 
\[
D=(-1)^{l(l-1)/2}2^l
\prod_{a<b}
\la_{i_a,j_b}\la_{j_a,i_b}\la_{i_a,i_b}\la_{j_a,j_b}
\prod_{a,c}\la_{i_a,k_c}\la_{j_a,k_c}, 
\]
${\bf s}=
s^{(2)}_{(i_1,\overline{i_1}),(j_1,\overline{j_1})}
\cdots s^{(2)}_{(i_m,\overline{i_m}),(j_m,\overline{j_m})}
s^{(1)}_{k_1,\overline{k_1}}
\cdots s^{(1)}_{k_l,\overline{k_l}}$ 
is the product of singlet vectors, 
and $\tau_a$'s are transfer matrices 
\be
&&\tau_a
=\frac{1}{\la_{i_a,j_a}(\la_{i_a,j_a}^2-1)}
\Tr_{\la_{i_a,j_a}}
\Bigl(T_{n-2a+2}\bigl(\frac{\la_{i_a}+\la_{j_a}}{2}\bigr)\Bigr),
\\
&&
T_{n-2a+2}(\la)=
L_{\overline{\sigma(2a)}}
(\la-\la_{\sigma(2a)}-1)
\cdots
L_{\overline{\sigma(n)}}(\la-\la_{\sigma(n)}-1)
\\
&&\quad \times
L_{\sigma(n)}(\la-\la_{\sigma(n)})\cdots 
L_{\sigma(2a)}(\la-\la_{\sigma(2a)}).
\en
\medskip

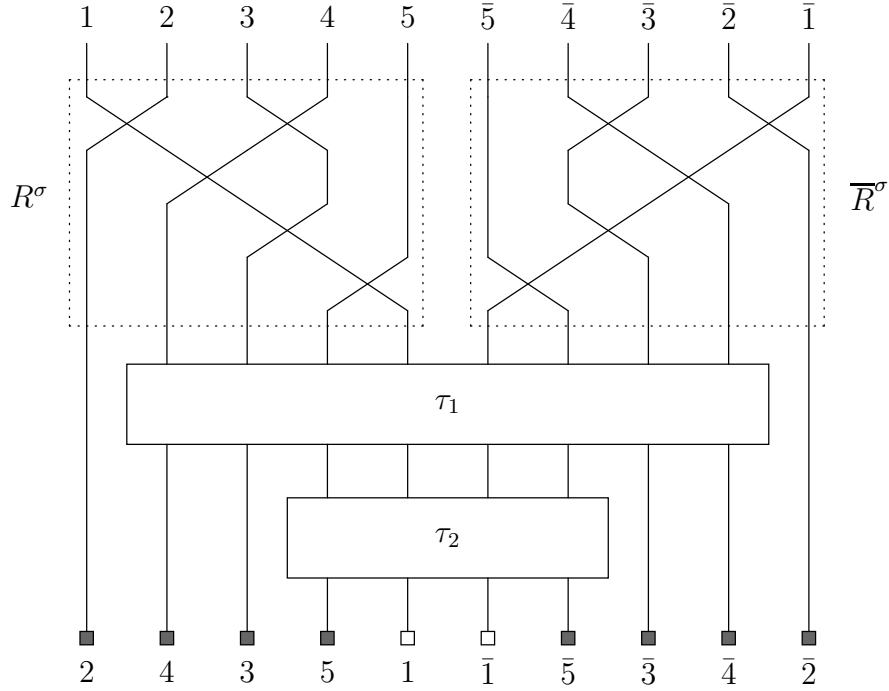
\begin{figure}
\unitlength 0.1in
\begin{picture}( 54.0000, 34.3000)( -1.4000,-39.0500)
%
\special{pn 8}%
\special{pa 700 700}%
\special{pa 700 980}%
\special{fp}%
%
\special{pn 8}%
\special{pa 1120 700}%
\special{pa 1120 980}%
\special{fp}%
\special{pa 1120 980}%
\special{pa 1120 980}%
\special{fp}%
%
\special{pn 8}%
\special{pa 1540 700}%
\special{pa 1540 980}%
\special{fp}%
%
\special{pn 8}%
\special{pa 1960 700}%
\special{pa 1960 980}%
\special{fp}%
%
\special{pn 8}%
\special{pa 2380 700}%
\special{pa 2380 980}%
\special{fp}%
%
\special{pn 8}%
\special{pa 2800 700}%
\special{pa 2800 980}%
\special{fp}%
%
\special{pn 8}%
\special{pa 3220 700}%
\special{pa 3220 980}%
\special{fp}%
%
\special{pn 8}%
\special{pa 3640 700}%
\special{pa 3640 980}%
\special{fp}%
%
\special{pn 8}%
\special{pa 4060 700}%
\special{pa 4060 980}%
\special{fp}%
%
\special{pn 8}%
\special{pa 4480 700}%
\special{pa 4480 980}%
\special{fp}%
%
\special{pn 8}%
\special{pa 700 980}%
\special{pa 1120 1260}%
\special{fp}%
%
\special{pn 8}%
\special{pa 1120 980}%
\special{pa 700 1260}%
\special{fp}%
%
\special{pn 8}%
\special{pa 1540 980}%
\special{pa 1960 1260}%
\special{fp}%
%
\special{pn 8}%
\special{pa 1960 980}%
\special{pa 1540 1260}%
\special{fp}%
%
\special{pn 8}%
\special{pa 1120 1260}%
\special{pa 2380 2100}%
\special{fp}%
%
\special{pn 8}%
\special{pa 1540 1260}%
\special{pa 1120 1540}%
\special{fp}%
%
\special{pn 8}%
\special{pa 1960 1260}%
\special{pa 1960 1540}%
\special{fp}%
%
\special{pn 8}%
\special{pa 1960 1540}%
\special{pa 1540 1820}%
\special{fp}%
%
\special{pn 8}%
\special{pa 2380 980}%
\special{pa 2380 1820}%
\special{fp}%
%
\special{pn 8}%
\special{pa 2380 1820}%
\special{pa 1960 2100}%
\special{fp}%
%
\special{pn 8}%
\special{pa 4480 980}%
\special{pa 2800 2100}%
\special{fp}%
%
\special{pn 8}%
\special{pa 4060 980}%
\special{pa 4480 1260}%
\special{fp}%
%
\special{pn 8}%
\special{pa 3640 980}%
\special{pa 3220 1260}%
\special{fp}%
%
\special{pn 8}%
\special{pa 3220 980}%
\special{pa 3640 1260}%
\special{fp}%
%
\special{pn 8}%
\special{pa 3640 1260}%
\special{pa 4060 1540}%
\special{fp}%
%
\special{pn 8}%
\special{pa 3220 1260}%
\special{pa 3220 1540}%
\special{fp}%
%
\special{pn 8}%
\special{pa 3220 1540}%
\special{pa 3640 1820}%
\special{fp}%
%
\special{pn 8}%
\special{pa 2800 980}%
\special{pa 2800 1820}%
\special{fp}%
%
\special{pn 8}%
\special{pa 2800 1820}%
\special{pa 3220 2100}%
\special{fp}%
%
\special{pn 8}%
\special{pa 2380 2100}%
\special{pa 2380 2380}%
\special{fp}%
%
\special{pn 8}%
\special{pa 1960 2100}%
\special{pa 1960 2380}%
\special{fp}%
%
\special{pn 8}%
\special{pa 1540 1820}%
\special{pa 1540 2380}%
\special{fp}%
%
\special{pn 8}%
\special{pa 1120 1540}%
\special{pa 1120 2380}%
\special{fp}%
%
\special{pn 8}%
\special{pa 2800 2100}%
\special{pa 2800 2380}%
\special{fp}%
%
\special{pn 8}%
\special{pa 3220 2100}%
\special{pa 3220 2380}%
\special{fp}%
%
\special{pn 8}%
\special{pa 3640 1820}%
\special{pa 3640 2380}%
\special{fp}%
%
\special{pn 8}%
\special{pa 4060 1540}%
\special{pa 4060 2380}%
\special{fp}%
%
\special{pn 8}%
\special{pa 910 2380}%
\special{pa 4270 2380}%
\special{pa 4270 2800}%
\special{pa 910 2800}%
\special{pa 910 2380}%
\special{fp}%
%
\special{pn 8}%
\special{pa 1960 2800}%
\special{pa 1960 3080}%
\special{fp}%
%
\special{pn 8}%
\special{pa 2380 2800}%
\special{pa 2380 3080}%
\special{fp}%
%
\special{pn 8}%
\special{pa 2800 2800}%
\special{pa 2800 3080}%
\special{fp}%
%
\special{pn 8}%
\special{pa 3220 2800}%
\special{pa 3220 3080}%
\special{fp}%
%
\special{pn 8}%
\special{pa 1750 3080}%
\special{pa 3430 3080}%
\special{pa 3430 3500}%
\special{pa 1750 3500}%
\special{pa 1750 3080}%
\special{fp}%
%
\special{pn 8}%
\special{pa 1960 3500}%
\special{pa 1960 3780}%
\special{fp}%
%
\special{pn 8}%
\special{pa 2380 3500}%
\special{pa 2380 3780}%
\special{fp}%
%
\special{pn 8}%
\special{pa 2800 3500}%
\special{pa 2800 3780}%
\special{fp}%
%
\special{pn 8}%
\special{pa 3220 3500}%
\special{pa 3220 3780}%
\special{fp}%
%
\special{pn 8}%
\special{pa 1540 2800}%
\special{pa 1540 3780}%
\special{fp}%
%
\special{pn 8}%
\special{pa 1120 2800}%
\special{pa 1120 3780}%
\special{fp}%
%
\special{pn 8}%
\special{pa 3640 2800}%
\special{pa 3640 3780}%
\special{fp}%
%
\special{pn 8}%
\special{pa 4060 2800}%
\special{pa 4060 3780}%
\special{fp}%
%
\special{pn 8}%
\special{pa 700 1260}%
\special{pa 700 3780}%
\special{fp}%
%
\special{pn 8}%
\special{pa 4480 1260}%
\special{pa 4480 3780}%
\special{fp}%
\put(7.0000,-5.6000){\makebox(0,0){$1$}}%
\put(11.2000,-5.6000){\makebox(0,0){$2$}}%
\put(15.4000,-5.6000){\makebox(0,0){$3$}}%
\put(19.6000,-5.6000){\makebox(0,0){$4$}}%
\put(23.8000,-5.6000){\makebox(0,0){$5$}}%
\put(28.0000,-5.6000){\makebox(0,0){$\bar{5}$}}%
\put(32.2000,-5.6000){\makebox(0,0){$\bar{4}$}}%
\put(36.4000,-5.6000){\makebox(0,0){$\bar{3}$}}%
\put(40.6000,-5.6000){\makebox(0,0){$\bar{2}$}}%
\put(44.8000,-5.6000){\makebox(0,0){$\bar{1}$}}%
\put(23.8000,-39.9000){\makebox(0,0){$1$}}%
\put(19.6000,-39.9000){\makebox(0,0){$5$}}%
\put(15.4000,-39.9000){\makebox(0,0){$3$}}%
\put(11.2000,-39.9000){\makebox(0,0){$4$}}%
\put(7.0000,-39.9000){\makebox(0,0){$2$}}%
\put(28.0000,-39.9000){\makebox(0,0){$\bar{1}$}}%
\put(32.2000,-39.9000){\makebox(0,0){$\bar{5}$}}%
\put(36.4000,-39.9000){\makebox(0,0){$\bar{3}$}}%
\put(40.6000,-39.9000){\makebox(0,0){$\bar{4}$}}%
\put(44.8000,-39.9000){\makebox(0,0){$\bar{2}$}}%
\put(25.9000,-25.9000){\makebox(0,0){$\tau_{1}$}}%
\put(25.9000,-32.9000){\makebox(0,0){$\tau_{2}$}}%
%
\special{pn 8}%
\special{pa 2346 3780}%
\special{pa 2416 3780}%
\special{pa 2416 3850}%
\special{pa 2346 3850}%
\special{pa 2346 3780}%
\special{fp}%
%
\special{pn 8}%
\special{pa 2766 3780}%
\special{pa 2836 3780}%
\special{pa 2836 3850}%
\special{pa 2766 3850}%
\special{pa 2766 3780}%
\special{fp}%
%
\special{pn 8}%
\special{sh 0.600}%
\special{pa 1926 3780}%
\special{pa 1996 3780}%
\special{pa 1996 3850}%
\special{pa 1926 3850}%
\special{pa 1926 3780}%
\special{fp}%
%
\special{pn 8}%
\special{sh 0.600}%
\special{pa 1506 3780}%
\special{pa 1576 3780}%
\special{pa 1576 3850}%
\special{pa 1506 3850}%
\special{pa 1506 3780}%
\special{fp}%
%
\special{pn 8}%
\special{sh 0.600}%
\special{pa 1086 3780}%
\special{pa 1156 3780}%
\special{pa 1156 3850}%
\special{pa 1086 3850}%
\special{pa 1086 3780}%
\special{fp}%
%
\special{pn 8}%
\special{sh 0.600}%
\special{pa 666 3780}%
\special{pa 736 3780}%
\special{pa 736 3850}%
\special{pa 666 3850}%
\special{pa 666 3780}%
\special{fp}%
%
\special{pn 8}%
\special{sh 0.600}%
\special{pa 3186 3780}%
\special{pa 3256 3780}%
\special{pa 3256 3850}%
\special{pa 3186 3850}%
\special{pa 3186 3780}%
\special{fp}%
%
\special{pn 8}%
\special{sh 0.600}%
\special{pa 3606 3780}%
\special{pa 3676 3780}%
\special{pa 3676 3850}%
\special{pa 3606 3850}%
\special{pa 3606 3780}%
\special{fp}%
%
\special{pn 8}%
\special{sh 0.600}%
\special{pa 4026 3780}%
\special{pa 4096 3780}%
\special{pa 4096 3850}%
\special{pa 4026 3850}%
\special{pa 4026 3780}%
\special{fp}%
%
\special{pn 8}%
\special{sh 0.600}%
\special{pa 4446 3780}%
\special{pa 4516 3780}%
\special{pa 4516 3850}%
\special{pa 4446 3850}%
\special{pa 4446 3780}%
\special{fp}%
%
\special{pn 8}%
\special{pa 610 890}%
\special{pa 2460 890}%
\special{pa 2460 2180}%
\special{pa 610 2180}%
\special{pa 610 890}%
\special{dt 0.045}%
%
\special{pn 8}%
\special{pa 2710 890}%
\special{pa 4560 890}%
\special{pa 4560 2180}%
\special{pa 2710 2180}%
\special{pa 2710 890}%
\special{dt 0.045}%
\put(4.0000,-15.0000){\makebox(0,0){$R^{\sigma}$}}%
\put(48.0000,-15.0000){\makebox(0,0){$\overline{R}^{\sigma}$}}%
\end{picture}%

\caption{The case $n=5$, $I=(23)$, $J=(45)$.}
\end{figure}

\medskip

\subsection{Examples}\label{subsec:2.6}
Let us write down the recursion relation in simple cases.

First let us take $n=2$. Noting that
\be
\Tr_x(AB)=\frac{1}{6}x(x^2-1)(A|B)\qquad (A,B\in \slt)
\en
and using the initial condition \eqref{eq:initial}, 
we find 
\be
X^{[1,2]}(\la_1,\la_2)=\frac{1}{3}s^{(2)}_{(1,\bar{1}),(2,\bar{2})}.
\en
Along with the initial condition \eqref{eq:initial}, 
the recursion formula gives 
\be
h_2(\la_1,\la_2)=\frac{1}{3}\omega(\la_{1,2})
s^{(2)}_{(1,\bar{1}),(2,\bar{2})}
-\frac{1}{4}s^{(1)}_{1,\bar{1}}s^{(1)}_{2,\bar{2}}.
\en

Next consider the case $n=3$. We have
\be
\Tr_x(ABC)=\frac{1}{12}x(x^2-1)([A,B]|C)\qquad (A,B,C\in \slt).
\en
Take a basis of the $\slt$-invariants $(V^{\otimes 6})^{\slt}$ as follows:
\be
&&u_0=s^{(1)}_{1,\bar{1}}s^{(1)}_{2,\bar{2}}s^{(1)}_{3,\bar{3}},
\\
&&u_i=s^{(1)}_{i,\bar{i}}s^{(2)}_{(j,\bar{j}),(k,\bar{k})},
\quad (\mbox{$i,j,k=1,2,3$ are distinct}),
\\
&&u_4=u,
\en
where $u$ is the unique 
$\slt$-invariant vector in 
$V^{(2)}_{(1,\bar{1})}\otimes V^{(2)}_{(2,\bar{2})}
\otimes V^{(2)}_{(3,\bar{3})}$
with coefficient $1$ in the component
$v^{(2)}_0\otimes v^{(2)}_1\otimes v^{(2)}_2$.

After some calculation we find 
\be
&&
X^{[1,2]}(\la_1,\la_2,\la_3)s^{(1)}_{3,\bar{3}}
=
\frac{1}{3}u_3
+\frac{1}{3\la_{1,3}\la_{2,3}}u_2
-\frac{\la_{1,2}}{6\la_{1,3}\la_{2,3}}u_4,
\en
which gives
\be
h_3(\la_1,\la_2,\la_3)
&=&
\omega(\la_{1,2})
\left(\frac{1}{6}u_3
+\frac{1}{6\la_{1,3}\la_{2,3}}u_2
-\frac{\la_{1,2}}{12\la_{1,3}\la_{2,3}}u_4\right)
\\
&+&
\omega(\la_{1,3})
\left(\frac{1}{6}
\Bigl(1-\frac{1}{\la_{1,2}\la_{2,3}}\Bigr)u_2
+\frac{\la_{1,2}-\la_{2,3}}{12\la_{1,2}\la_{2,3}}u_4
\right)
\\
&+&
\omega(\la_{2,3})
\left(\frac{1}{6}u_1
+\frac{1}{6\la_{1,2}\la_{1,3}}u_2
+\frac{\la_{2,3}}{12\la_{1,2}\la_{1,3}}u_4\right)
\\
&-&\frac{1}{8}u_0.
\en

\section{Derivation from the qKZ equation}\label{sec:4}
The purpose of this section is to 
give a derivation of Theorem \ref{thm:recursion}--\ref{prop:ansatz2}.

\subsection{Properties of $X^{[i,j]}(\la_1,\cdots,\la_n)$}\label{subsec:4.1}

In this subsection, we study the properties 
of the rational function $X^{[i,j]}(\la_1,\cdots,\la_n)$.

We begin by showing an identity of $L$ operators.
\begin{lem}\label{LemL} 
The following equality holds. 
\begin{equation}\label{EQUAL}
\varpi_\la
\left(L_2(\frac{\la}{2})L_{\bar 2}(\frac{\la}{2}-1)\right)
s^{(2)}_{(1\bar1),(2\bar 2)}
=
\frac{1}{\la+2}r^{(2,1)}_{(1,\bar{1}),\bar{2}}(\la+\frac{1}{2})
\varpi_\la\left(L^{(2)}_{(2,\bar{2})}\(\frac{\la+1}{2}\)\right)
s^{(2)}_{(1,\bar{1}),(2,\bar{2})}.
\end{equation}
\end{lem}
\begin{proof}
We show (\ref{EQUAL}) by projecting it to 
the symmetric and skew-symmetric
subspaces of $V_{2}\otimes V_{\bar{2}}$, separately. 
To simplify the notation, in what follows 
we shall write $A\sim B$ to indicate the equality 
$\varpi_{\la}(A)=\varpi_{\la}(B)$. 
Note that
\begin{equation}
C\sim\frac{\la^2-1}{2}.
\end{equation}

{}From (\ref{eq:Lop2}) we have
\be
\P^+_{2,\bar2}
L_2(\frac{\la}{2})L_{\bar 2}(\frac{\la}{2}-1)s^{(2)}_{(1\bar1),(2\bar 2)}
=L^{(2)}_{(2\bar2)}(\frac{\la-1}{2})s^{(2)}_{(1\bar1),(2\bar 2)}.
\en
Hence, on the symmetric subspace, 
\eqref{EQUAL} reduces to 
\begin{equation}\label{CHK}
L^{(2)}_{(2\bar2)}\left(\frac {\la-1}{2}\right)s^{(2)}_{(1\bar1),(2\bar 2)}
\sim
\frac{1}{\la+2}\left(\la+1+\frac{1}{2}\sum_a\(S_a\)_{(1\bar{1})}\(S^a\)_{(2\bar{2})}
\right)L^{(2)}_{(2\bar2)}\left(\frac{\la+1}{2}\right)s^{(2)}_{(1\bar1),(2\bar 2)}.
\end{equation}
To simplify the right hand side, notice the relation 
\begin{equation}\label{SPECT}
\frac{1}{2}\sum_a S_a\otimes S^a=P^{(2)}-3 K^{(2)},
\end{equation}
where 
$P^{(2)}\in{\rm End}_{\frak{sl}_2}(V^{(2)}\otimes V^{(2)})$ 
is the permutation operator and 
$K^{(2)}\in{\rm End}_{\frak{sl}_2}(V^{(2)}\otimes V^{(2)})$ is 
the projector onto the singlet subspace. 
More explicitly, we have
\[
K^{(2)}=\frac13s^{(2)}\otimes s^{(2)}\in (V^{(2)}\otimes V^{(2)})^{\otimes2}
\simeq
{\rm End}(V^{(2)}\otimes V^{(2)}),
\]
where we made 
the identification through the invariant bilinear form
$(\cdot,\cdot):V^{(2)}\otimes V^{(2)}\rightarrow\C$
normalized as $(v^{(2)}_0,v^{(2)}_2)=1$.

By using (\ref{SPECT}), the right hand side of (\ref{CHK}) can be rewritten as 
\begin{eqnarray*}
\frac{1}{\la+2}\((\la+1)L^{(2)}_{(2\bar2)}\left(\frac{\la+1}{2}\right)
+L^{(2)}_{(2\bar2)}\left(-\frac{\la+3}{2}\right)
-
\tr_{(2\bar{2})}\(L^{(2)}_{(2\bar2)}\left(\frac{\la+1}{2}\right)\)
\)s^{(2)}_{(1\bar1),(2\bar 2)}. 
\nn
\end{eqnarray*}
In the second term, we used the crossing 
symmetry (\ref{eq:crossing}). 
The trace is evaluated as 
\begin{equation}
\tr _{(2\bar{2})}\(L^{(2)}_{(2\bar2)}\left(\frac{\la+1}{2}\right)\)\sim (\la+1)(\la+2).
\end{equation}
Now one can check (\ref{CHK}) directly by using (\ref{L^2}).

Next let us project to the singlet subspace. 
The right hand side becomes
\begin{equation}
\P^-_{2,\bar2}\hbox{(RHS)}
=\frac1{2(\la+2)}\sum_a(S_a)_{(1,\bar1)}
\((S^a)_{\bar2}-(S^a)_2\)L^{(2)}_{(2,\bar2)}
\bigl(\frac{\la+1}{2}\bigr)s^{(2)}_{(1\bar1),(2\bar 2)}.
\end{equation}
Using the matrix representation
\begin{eqnarray*}
&&L^{(2)}_{(2\bar2)}\left(\frac{\lambda+1}2\right)
\sim\begin{pmatrix}
\frac{(H+\la+1)(H+\la+3)}4&(H+\la+3)F&F^2\\
\frac{E(H+\la+3)}{2}&\frac{(\la+1)^2-H^2}{2}&-\frac{F(H-\la-3)}{2}\\
E^2&-(H-\la-3)E&\frac{(H-\la-1)(H-\la-3)}{4}\\
\end{pmatrix},
\end{eqnarray*}
we obtain 
\[
\P^-_{2,\bar2}\hbox{(RHS)}\sim\(-F(v^{(2)}_0)_{(1\bar1)}
+\frac12H(v^{(2)}_1)_{(1\bar1)}+E(v^{(2)}_2)_{(1\bar1)}\)s^{(1)}_{2,\bar2}.
\]
The left hand side becomes 
\begin{eqnarray*}
&&\P^-_{2,\bar2}\hbox{(LHS)}\\
&&=\left(\P^-_{2,\bar2}
L_{\bar 2}\bigl(\frac{\la-2}{2})L_2(\frac{\la}{2}\bigr)
+\P^-_{2\bar2}
\left[L_2\bigl(\frac{\la}{2}\bigr),
L_{\bar 2}\bigl(\frac{\la-2}{2}\bigr)
\right]\right)s^{(2)}_{(1\bar1),(2\bar 2)}. 
\end{eqnarray*}
The first term is zero because of (\ref{SINGL}). 
Therefore, we have 
\begin{equation}
\P^-_{2,\bar2}\hbox{(LHS)}
\sim
\P^-_{2,\bar2}\sum_{a,b}[S_a,S_b](S^a)_2(S^b)_{\bar2}
s^{(2)}_{(1\bar1),(2\bar 2)}.
\end{equation}
Now the equality 
$\P^-_{2, \bar2}\hbox{(LHS)}\sim\P^-_{2,\bar2}\hbox{(RHS)}$
is easy to see. 
\end{proof}

We have the following symmetry of $X^{[1,2]}$.
\begin{lem}\label{lem:X2sym}
The function $X^{[1,2]}(\la _1,\cdots ,\la _n)$ possesses 
the symmetry property 
\begin{align}
&R_{2,1}(\la _{2,1})R_{\bar1,\bar2}(\la_{1,2})
X^{[1,2]}(\la_2,\la _1, \cdots,\la_n)_{2,1,\cdots, n,\bar n,\cdots,\bar1,\bar2}
\label{SymX}\\
&=
X^{[1,2]}(\la_1,\la _2, \cdots,\la_n)_{1,2,\cdots n,\bar n,\cdots,\bar2,\bar1}\,.
\nn
\end{align}
\end{lem}
\begin{proof}
Recall the definition
\begin{eqnarray}
&& 
X^{[1,2]}(\la_1,\cdots,\la_n)
\label{eq:recall_X12} \\ 
&&{}=
\frac1{\lambda_{1,2}(\la _{1,2}^2-1)
\prod_{p\not=1,2}\lambda_{1,p}\lambda_{2,p}}
\Tr_{\lambda_{1,2}}\(T^{[1]}\left(
\frac{\lambda_1+\lambda_2}2\right)\)
s^{(2)}_{(1,\bar1),(2,\bar2)},
\nonumber 
\end{eqnarray}
where
\bea
{}\quad 
T^{[1]}(\lambda)=
L_{\bar2}(\lambda-\lambda_2-1)\cdots 
L_{\bar n}(\lambda-\lambda_n-1)
L_n(\lambda-\lambda_n)\cdots L_2(\lambda-\lambda_2).
\label{eq:recall_T}
\ena
{}From this formula, we see that \eqref{SymX}
is proven once we verify the relation 
\begin{align}
&\varpi_\lambda
\(R_{2,1}(-\la)R_{\bar1,\bar2}(\la)L_{1}\(-\frac{\la }2\)
L_{\bar1}\(-\frac{\la } 2-1\)s^{(2)}_{(1\bar1),(2\bar2)}\)
\label{RRL}\\
&=\varpi_\lambda
\(L_{2}\(\frac{\la}2\)L_{\bar2}\(\frac{\la}2-1\)s^{(2)}_{(1\bar1),(2\bar2)}\),
\nn
\end{align}
where we denoted $\la _{1,2}$ by $\la$. 
As before, let us 
write $A\sim B$ to mean 
$\varpi_\lambda(A)=\varpi_\lambda(B)$. 
Using Lemma \ref{LemL}, (\ref{RRL}) can be rewritten as 
\begin{align}
&R_{2,1}(-\la )R_{\bar1,\bar2}(\la)
\frac 1 {\la -2}r^{(2,1)}_{(2,\bar{2}),\bar{1}}
\(-\la +\frac{1}{2}\)
L^{(2)}_{(1,\bar{1})}\(\frac{-\la +1}{2}\)
s^{(2)}_{(1\bar1),(2\bar2)}\nn\\&\sim
-\frac1{\la +2}r^{(2,1)}_{(1,\bar{1}),\bar{2}}\(\la+\frac{1}{2}\)
L^{(2)}_{(2,\bar{2})}\(\frac{\la +1}{2}\)
s^{(2)}_{(1\bar1),(2\bar2)}\,,\nn
\end{align}
which is further equivalent to
\bea
&&r^{(2,2)}_{(1\bar{1}),(2\bar{2})}(\la)
L^{(2)}_{(2,\bar{2})}\(\frac{\la +1}{2}\)s^{(2)}_{(1\bar1),(2\bar2)}
\label{eq:RRL2}\\
&&
\sim
(\la +2)(\la +1)L^{(2)}_{(1,\bar{1})}\(\frac{-\la +1}{2}\)
s^{(2)}_{(1\bar1),(2\bar2)}\,
\nn
\ena
by using \eqref{L4} and \eqref{L6}.
The proof of \eqref{eq:RRL2} is similar to that of \eqref{CHK}.
We use the spectral decomposition
\bea
r^{(2,2)}(\la)=\la(\la +1)I+2(\la +1)P^{(2)}-6\la K^{(2)}
\label{eq:spectral}
\ena
which follows from \eqref{SPECT} and the equality 
$(1/2)\sum_{a,b} {(S_a S_b)}_{(1,\bar 1)} 
\otimes {(S^a S^b)}_{(2,\bar 2)}= 2 - 6 K^{(2)}$. 
Here $P^{(2)}, K^{(2)}$ have the same meaning as in \eqref{SPECT}.

Substituting \eqref{eq:spectral} into the left hand side of 
\eqref{eq:RRL2} we obtain 
\begin{align}
r^{(2,2)}_{(1\bar{1}),(2\bar{2})}(\la)&
L^{(2)}_{(2,\bar{2})}\(\frac{\la +1}{2}\)
s^{(2)}_{(1\bar1),(2\bar2)}=
\(\la (\la +1)L^{(2)}_{(1,\bar{1})}\(-1-\frac{\la +1}{2}\)+\right.\nn\\&+\left.
2(\la+1)L^{(2)}_{(1,\bar{1})}\(\frac{\la +1}{2}\)-2\la(\la+1)(\la+2)
\)s^{(2)}_{(1\bar1),(2\bar2)}.\nn
\end{align}
The equality \eqref{eq:RRL2} can now 
be verified by a straightforward calculation. 
\end{proof}

\begin{lem}\label{lem:trans}
$X^{[i,j]}$ obey the transformation rules \eqref{TRANS}.
\end{lem}
\begin{proof}
Except for the case $k=i=j-1$, 
these are immediate consequences of the definition and 
the Yang-Baxter relation.
The non-trivial case $k=i=j-1$ follows from Lemma 
\ref{lem:X2sym} and the Yang-Baxter relation.  
\end{proof}

The next lemma concerns the pole structure of $X^{[i,j]}$.
\begin{lem}\label{lem:pole_of_X}
$\prod_{p(\neq i,j)}\la_{i,p}\la_{j,p}\cdot X^{[i,j]}(\la_1,\cdots,\la_n)$ 
is a polynomial. 
\end{lem}
\begin{proof}
First let us consider $X^{[1,2]}$ {}
given by \eqref{eq:recall_X12}, \eqref{eq:recall_T}.
We have to prove that 
the poles at $\la_1=\la_2$, $\la _1=\la _2\pm 1$ are spurious.
Because of the property \eqref{eq:tr3} of $\Tr_x$, 
it is enough to show that 
$\varepsilon\left(T^{[1]}\bigl((\la_1+\la_2)/2\bigr)\right)$
is divisible by $\la_{1,2}^2-1$. 
This is indeed the case, since
$\varepsilon\left(L_{\bar{2}}\bigl(\la_{1,2}/2-1\bigr)\right)
=(\la_{1,2}-1)/2$ and 
$\varepsilon\left(L_2\bigl(\la_{1,2}/2\bigr)\right)=(\la_{1,2}+1)/2$.

Next consider the case $i=1$ with $j$ general.
We use the relation between $X^{[1,j]}$ and $X^{[1,2]}$, 
\bea
&&X^{[1,j]}(\la_1,\cdots,\la_n)
=R_{j,j-1}(\lambda_{j,j-1})\cdots R_{j,2}(\lambda_{j,2})
\label{eq:X2Xj}
\\
&\times&
R_{\overline{j-1},\bar{j}}(\lambda_{j-1,j})\cdots 
\cdots R_{\bar{2},\bar{j}}(\lambda_{2,j})
\nn
\\
&\times&
X^{[1,2]}(\la_1,\la_j,\la_2,\cdots,\widehat{\la_j},\cdots,\la_n)
_{1,j,2,\cdots,\hat{j},\cdots,n,
\bar{n},\cdots,\hat{\bar{j}},\cdots,\bar{2},
\bar{j},\bar{1}}.
\nn
\ena 
We are to show that the poles $\la _{i}=\la _j \pm 1$ 
($2\le i<j$) 
contained in the $R$ matrices are in fact spurious. 
Taking the residue of $X^{[1,j]}$ at $\la _i=\la _j+1$ ($j>i$), 
we find the following fragment
\begin{eqnarray}
\Tr_{\lambda_{1,j}}\Bigr(T^{[1]}\left(\frac{\la_1+\la_j}2\right)\Bigl)
R_{j,j-1}(\lambda_{j,j-1})\cdots R_{j,i+1}(\lambda_{j,i+1})
\mathcal{P}^-_{i,j}\,.\nn
\end{eqnarray}
Move the $R$-matrices to the 
left by using the Yang-Baxter relation, which
permutes the $L$-operators in $T^{[1]}$. 
This will bring the $L_j$ next to $L_i$. 
Thus we find 
\begin{eqnarray}
&L_j\(\frac{\la_{1,j}}{2}\)L_i\(\frac{\la_{1,j}}{2}-1\)
\mathcal{P}^-_{i,j}=0,\nn
\end{eqnarray}
where we used 
the equation for the quantum determinant (\ref{SINGL}) 
and the fact that the 
`dimension' of the auxiliary space equals to $\la_{1,j}$.
The pole at $\la _i=\la _j-1$ ($j>i$) can be treated similarly.

For general $X^{[i,j]}$, we use the symmetry \eqref{SymX}
to obtain the two representations
\be
&&X^{[i,j]}(\la_1,\cdots,\la_n)
=R_{i,i-1}\cdots R_{i,1}R_{\overline{i-1},\bar{i}}\cdots R_{\bar{1},\bar{i}}
\\
&&\quad\times 
X^{[1,j]}(\la_i,\la_1,\cdots,\widehat{\la_i},\cdots,\la_n)
_{i,1,\cdots,\hat{i},\cdots,n,\bar{n},\cdots,\hat{\bar{i}},\cdots,
\bar{1},\bar{i}}
\\
&&=
R_{j,j-1}\cdots R_{j,1}R_{i+1,i}\cdots R_{j-1,i}
R_{\overline{j-1},\bar{j}}\cdots R_{\bar{1},\bar{j}}
R_{\bar{i},\overline{i+1}}\cdots R_{\bar{i},\overline{j-1}}
\\
&&\times 
X^{[1,j]}(\la_j,\la_1,\cdots,\widehat{\la_i},\cdots,
\la_{j-1},\la_i,\la_{j+1},\cdots,\la_n)
_{j,1,\cdots,\hat{i},\cdots,j-1,i,j+1,\cdots,
n,\bar{n},\cdots,\overline{j+1},\bar{i},\overline{j-1},\cdots,
\hat{\bar{i}},\cdots,
\bar{1},\bar{j}}\,.
\en
where $R_{i,j}=R_{i,j}(\la_{i,j})$ and 
$R_{\bar{j},\bar{i}}=R_{\bar{j},\bar{i}}(\la_{j,i})$. 
{}From these expressions, 
it is clear that the poles 
arising from the $R$ matrices are spurious. 
\end{proof}

The following lemma shows that 
the function $X^{[i,j]}(\la_1,\cdots,\la_n)$ 
is regular at $\la_k=\infty$ 
when it is applied to a singlet vector. 
 
\begin{lem}\label{lem:degree_of_X}
{}For any vector
$v\in\bigl(W^{[i,j]}\bigr)^{\slt}$ and $k=1,\cdots,n$, we have 
\be
X^{[i,j]}(\la_1,\cdots,\la_n)\cdot v =O(1)
\qquad (\la_k\to \infty).
\en
\end{lem}
\begin{proof}
Since $R_{i,j}(\la_{i,j})R_{\bar{j},\bar{i}}(\la_{j,i})$ is holomorphic 
and invertible at $\la_{i,j}=\infty$, it is enough to consider the 
case $(i,j)=(1,2)$. 
For $k\ge 3$, the assertion is clear from the definition \eqref{eq:Xij}.
We will prove the case $k=1$.  
The case $k=2$ reduces to this case 
by the symmetry \eqref{SymX}. 

As before, we write $A\sim B$ for $\varpi_x(A)=\varpi_x(B)$.
We show the following statement: {}For any $\mu_1,\cdots,\mu_{2N}\in \C$ and  
$u\in \bigl(V^{\otimes 2N}\bigr)^{\slt}$, there exist $c_{pqrs}\in\C$ such that
\bea
&& L_1(\mu_1+x/2)\cdots L_{2N}(\mu_{2N}+x/2)u\label{eq:degree<n}\\
&&\sim\left(\sum_{p+q+r+s\le N}c_{pqrs}H^pE^qF^rx^s\right)u.\nn
\ena
The assertion then follows from the property \eqref{eq:deg_of_Tr} of $\Tr_x$
by taking $N=n-2$.  

We may assume that the $\mu_j$'s are mutually distinct. 
It is sufficient to prove \eqref{eq:degree<n} for $u=u_\sigma=\sigma u_1$, 
where $u_1=s^{(1)}_{1,2}s^{(1)}_{3,4}\cdots s^{(1)}_{2N-1,2N}$ 
and $\sigma\in \mathfrak{S}_{2N}$. 
We use the induction on the length $\ell(\sigma)$. 
First consider the case $\sigma=\id$. 
Set $\Omega=\sum_{a=1}^3 S_a\otimes \pi^{(1)}(S^a)$. 
Since $\Omega_2s^{(1)}_{1,2}=-\Omega_1s^{(1)}_{1,2}$ and 
$\Omega^2+\Omega=\frac{1}{2}C\otimes \id\sim(x^2-1)/4$, we have
\be
&&
L_1(\mu_1+x/2) L_2(\mu_{2}+x/2)s^{(1)}_{1,2}
\\
&&\quad
\sim\left(\mu_1\mu_2+\frac{x+1}{2}(\mu_1+\mu_2)+(\mu_2-\mu_1+1)\Omega_1
+\frac{(x+1)^2}{4}-\frac{x^2-1}4\right)s^{(1)}_{1,2}.
\en
We see that the term $x^2$ cancels. Hence we have \eqref{eq:degree<n}
for $u=u_1$.

Suppose \eqref{eq:degree<n} is true for $u=u_\sigma$, and consider 
$\tau=(i,i+1)\sigma$ with $\ell(\tau)=\ell(\sigma)+1$. 
Using the Yang-Baxter relation, we have
\be
&&P_{i,i+1}r_{i,i+1}(\mu_i-\mu_{i+1})
L_{1}(\mu_1+\frac{x}{2})\cdots
L_{2N}(\mu_{2N}+\frac{x}{2})\cdot u_\sigma
\\
&&=
L_{1}(\mu_1+\frac{x}{2})\cdots
\cdots
L_{i}(\mu_{i+1}+\frac{x}{2})L_{i+1}(\mu_i+\frac{x}{2})
\\
&&\quad 
\cdots
L_{2N}(\mu_{2N}+\frac{x}{2})\cdot 
\bigl((\mu_i-\mu_{i+1})u_\tau+u_\sigma).
\en
By the induction hypothesis, the left hand side and the second term
in the right hand side have degree at most $N$ when they are projected
by $\varpi_x$. Therefore the statement is true also for $u=u_\tau$. 
\end{proof}

Using the properties of $X^{[i,j]}$, 
let us check that 
the pole structure of the recursion (\ref{eq:Recursion})
agrees with that of $h_n$ stated in (\ref{eq:meromorphy}). 
Namely, we have
\begin{prop}\label{ANALYTICITY}
Assume that the $h_{n-1}$ and $h_{n-2}$ satisfy the 
analyticity property
as stated in \eqref{eq:meromorphy}. Then the same is true for
$h_n$ given by \eqref{eq:Recursion}.
\end{prop}
\begin{proof}
We have shown in Lemma \ref{lem:pole_of_X} that in the right hand side
of \eqref{eq:Recursion},
the only possible poles other than $\la_{i,j}\in\Z\backslash\{0,\pm 1\}$
are $\la _i =\la _j$.
We rewrite the recursion as follows:
\bea
&&h_n(\la_1,\cdots,\la_n)_{1,\cdots,n,\bar{n},\cdots,\bar{1}}\nn\\
&=& \sum_{j=2}^n \frac1{\prod\limits_{p\not=1,j}\lambda_{jp}
\prod_{i=2}^{j-1}(1-\lambda_{ij}^2)}\cdot
\frac1{2\pi i}\oint_{\mathcal{C}}
\frac{d\sigma}{\prod_{p=1}^n(\sigma -\la_p)}\cdot
\frac{\omega(\sigma-\lambda_j)}{(\sigma-\lambda_j)^2-1}\nn\\
&\times&
\Tr_{\sigma-\lambda_j}
\Bigr(
T^{[1]}\bigl(\frac{\sigma+\lambda_j}{2}\bigr)
\Bigl)
\cdot r_{j,j-1}(\lambda_{j,j-1})\cdots r_{j,2}(\lambda_{j,2})
r_{\overline{j-1},\bar j}(\lambda_{j-1,j})\cdots r_{\bar2\bar j}(\lambda_{2j})
\nn\\&\times&s^{(2)}_{(1\bar1),(j\bar j)}
h_{n-2}(\la_2,\cdots,\hat{\la_j},\cdots,\la_n)_
{2,\cdots,\hat{j},\cdots,n,\bar{n},\cdots,\hat{\bar{j}},\cdots,\bar{2}}\nn \\
&+&(-1)^{n-1}\frac{1}{2}s^{(1)}_{(1\bar{1})}\cdot
h_{n-1}(\la_2,\cdots,\la_n)_{2,\cdots,n,\bar{n},\cdots,\bar{2}}.\nn
\ena
Here the contour $\mathcal{C}$ goes around $\la _1,\cdots ,\la _n$. 
{}From this expression, we see that the poles at $\la_{1p}=0$ are spurious.

Consider the pole at $\la _k=\la_j$ with $k>j>1$. 
We have
\begin{eqnarray}
&&\res _{\la _k=\la _j}\,h_n(\la _1,\cdots ,\la _n)\nn\\
&&=\frac1{\prod\limits_{p\not=1,j,k}\lambda_{jp}}
\frac1{2\pi i}\oint\frac{d\sigma}{\prod_{p=1}^n(\sigma -\la_p)}\cdot
\frac{\omega(\sigma-\lambda_j)}{(\sigma-\lambda_j)^2-1}\Tr_{\sigma-\lambda_j}
\Bigl(T^{[1]}\bigl(\frac{\sigma+\lambda_j}{2}\bigr)\Bigr)
\nn\\
&&\{-
R_{j,j-1}(\lambda_{j,j-1})\cdots R_{j,2}(\lambda_{j,2})
R_{\overline{j-1},\bar j}(\la_{j-1,j})
\cdots R_{\bar2\bar j}(\lambda_{2j})\nn\\
&&\times s^{(2)}_{(1\bar1),(j\bar j)}
h_{n-2}(\la_2,\cdots,\hat{\la_j},\cdots,\la_n)_
{2,\cdots,\hat{j},\cdots,n,
\bar{n},\cdots,\hat{\bar{j}},\cdots,\bar{2}}\Bigl|_{\la_k=\la_j}
\nn \\
&&+R_{k,k-1}(\la _{j,k-1})\cdots R_{k,j+1}(\la _{j,j+1})
R_{\overline{k-1},\bar{k}}(\la_{k-1,j})\cdots R_{\overline{j+1},\bar{k}}
(\la _{j+1,j})P_{j,k}P_{\bar{j},\bar{k}}\nn\\
&&\times R_{k,j-1}(\lambda_{j,j-1})\cdots R_{k,2}(\lambda_{j,2})
R_{\overline{j-1},\bar k}(\la_{j-1,j})\cdots R_{\bar2\bar k}(\lambda_{2j})\nn\\
&&\times s^{(2)}_{(1\bar1),(k\bar k)}
h_{n-2}(\la_2,\cdots,\hat{\la_k},\cdots,\la_n)_
{2,\cdots,\hat{k},\cdots,n,\bar{n},\cdots,\hat{\bar{k}},\cdots,\bar{2}}\bigr\}
\nn
\end{eqnarray}
In the second term inside the braces, 
move $P_{j,k}P_{\bar{j},\bar{k}}$
to the right, then apply symmetry (\ref{eq:Rsymm}).
After that the second term cancels the first one.
\end{proof}

\subsection{Proof of the Ansatz}\label{subsec;4.2}

In this subsection we prove the Ansatz \eqref{eq:Ansatz}.
\medskip

\noindent{\it Proof of Theorem \ref{prop:ansatz1}.}\quad 
The assertion is clear for $n=0,1$. 
By induction, assume that it is true for $h_{n'}$ with $n'<n$. 
We use the recursion formula in the form 
\bea
&&h_n(\la_1,\cdots,\la_n)_{1,\cdots,n,\bar{n},\cdots,\bar{1}}
\label{eq:recur}
\\
&&=\sum_{j=2}^n\omega(\la_{1,j})X^{[1,j]}(\la_1,\cdots,\la_n)
h_{n-2}(\la_2,\cdots,\widehat{\la_j},\cdots,\la_n)
_{2,\cdots,\hat{j},\cdots,n,\bar{n},\cdots,\hat{\bar{j}},\cdots,\bar{2}}
\nn\\
&&+\sum_{2\le j,k\le n\atop j\neq k}\frac{\omega(\la_{k,j})}{\la_{k,1}}
{\rm res}_{\sigma=\la_k}X^{[1,j]}(\sigma,\la_2,\cdots,\la_n)
\nn\\
&&\qquad\quad \times
h_{n-2}(\la_2,\cdots,\widehat{\la_j},\cdots,\la_n)
_{2,\cdots,\hat{j},\cdots,n,\bar{n},\cdots,\hat{\bar{j}},\cdots,\bar{2}}
\nn\\
&&+(-1)^{n-1}\frac{1}{2}s^{(1)}_{1,\bar{1}}
h_{n-1}(\la_2,\cdots,\la_n)
_{2,\cdots,n,\bar{n},\cdots,\bar{2}}.
\nn
\ena
By the induction hypothesis, 
the right hand side becomes a linear combination of the form 
\be
\prod_{p=1}^m\omega(\la_{i_p}-\la_{j_p})\,f_{n,I,J}(\la_1,\cdots,\la_n)
\en
with some rational function $f_{n,I,J}(\la_1,\cdots,\la_n)$
and $1\le i_1\le\cdots\le i_m\le n$, 
$1\le i_p<j_p\le n$. 
Since the product of the $\omega(\la)$'s are linearly independent over the 
field $\C(\la_1,\cdots,\la_n)$ of rational functions, 
this representation is unique.  
If there is a term for which $i_1,\cdots,i_m,j_1,\cdots,j_m$ are not distinct, 
then the symmetry relation \eqref{eq:Rsymm} of $h_n$ 
entails that there is also a term such that the index $1$ appears 
more than once in $i_1,\cdots,i_m,j_1,\cdots,j_m$. 
However, terms with $i_1=1$ 
arise only from the first term of \eqref{eq:recur} 
and the corresponding indices are distinct. 
Therefore we have \eqref{eq:ansatz}. 
\qed
\medskip

\noindent{\it Proof of Theorem \ref{prop:ansatz2}.}\quad 

{}From the above proof, the recursion relations 
\eqref{eq:f5}, \eqref{eq:f6} for $f_{n,I,J}$ are clear. 
Since the rational coefficients in \eqref{eq:ansatz} 
are unique, the $\slt$-invariance \eqref{eq:f1} 
of $f_{n,I,J}$ follows from that of $h_n$. 
Similarly, the exchange symmetry \eqref{eq:f4} 
follows from the symmetry \eqref{eq:Rsymm} of $h_n$ 
by comparing the coefficients of the $\omega$'s. 
The regularity \eqref{eq:f2} at $\infty$ is a consequence of 
Lemma \ref{lem:degree_of_X} and the $\slt$-invariance \eqref{eq:f1}. 
Finally the pole structure \eqref{eq:f3} 
follows from the recursion relations \eqref{eq:f5}, \eqref{eq:f6} and 
Lemma \ref{lem:pole_of_X}.
\qed

\subsection{Calculation of the residues}\label{subsec:4.3}
In this subsection, we compute the residues of $h_n(\la_1,\cdots,\la_n)$
with respect to $\la_1$. 
Our goal is Proposition \ref{prop:residue} and 
\ref{prop:residue1}.
Since the calculation is long, we outline the proof before starting. 

\medskip
We calculate the residues of the meromorphic function
$h_n(\lambda_1,\ldots,\lambda_n)$ at its poles. 
The poles are located 
at the points of the form $\lambda_{1,j}=k\in\Z\backslash\{0,\pm1\}$.
Then, we show that the right hand side of (\ref{eq:Recursion}) has
the same residues at these poles:
\begin{eqnarray}
&&\quad\res_{\la_{1j}=k}h_n(\la_1,\cdots,\la_n)_{1\cdots n\bar n\cdots\bar1}
\label{eq:1j}\\
&&=\res_{\la_{1j}=k}\{\omega(\la_{1,j})X^{[1,j]}(\la_1,\cdots,\la_n)\}
h_{n-2}(\la_2,\cdots,\widehat{\la_j},\cdots,\la_n)
_{2,\cdots,\hat j,\cdots,n,\bar{n},\cdots,\hat{\bar j},\cdots,\bar{2}}.\nn
\end{eqnarray}
A remarkable fact is that the residues at infinitely many points
can be recursively described by using a single analytic function
$\omega(\la_{1,j})X^{[1,j]}(\la_1,\cdots,\la_n)$. 
Thus, we show that the difference between the left and right hand sides
is an entire function of $\lambda_1$. 
Showing that it vanishes at $\lambda_1\rightarrow\infty$, we obtain the equality.

The calculation of residues goes as follows. Suppose $j=2$.
Using (\ref{eq:rqKZ}) repeatedly, we can express
$h_n(\lambda_1-k-1,\lambda_2,\ldots,\lambda_n)$
in terms of 
several $A_1$ and $A_{\bar1}$ with
shifted arguments acting on $h_n(\lambda_1-1,\lambda_2,\ldots,\lambda_n)$.
Using this expression, one can calculate the residue of
$h_n(\lambda_1,\lambda_2,\ldots,\lambda_n)$ at the pole
$\lambda_1=\lambda_2-k-1$. If $k=2$, for example, we have
\begin{eqnarray*}
&&{\rm res}_{\lambda_1=\lambda_2-3}h_n(\lambda_1,\lambda_2,\ldots,\lambda_n)\\
&=&{\rm res}_{\lambda_1=\lambda_2}
h_n(\lambda_1-3,\lambda_2,\ldots,\lambda_n)\\
&=&\res_{\lambda_1=\lambda_2}
\{A_{\bar1}(\lambda_1-2,\ldots)A_1(\lambda_1-1,\ldots)
h_n(\lambda_1-1,\ldots)\}
\end{eqnarray*}
Because of (\ref{eq:meromorphy}), the pole at $\lambda_1=\lambda_2$
in the last expression comes only from one of the $R$ matrices in
$A_1(\lambda_1-1,\ldots)$, i.e., $R_{1,2}(\lambda_{1,2}-1)$.
Its residue at $\lambda_{1,2}=0$ is proportional to the projector
$\P^-_{1,2}$, therefore one can use (\ref{eq:n_to_n-2}).
In this way, for any $k$, we get a certain expression for
${\rm res}_{\lambda_1=\lambda_2-k-1}h_n(\lambda_1,\ldots,\lambda_n)$;
a bunch of $R$ matrices acting on
$s^{(1)}_{\g2}s^{(1)}_{\og\overline{2}}h_{n-2}(\lambda_3,\ldots,\lambda_n)$,
where $\gamma=1$ ($k$: even) or $\bar1$ ($k$: odd).

Next we rewrite the product of $R$ matrices by using the Yang-Baxter relation.
They act on the spaces $V_\alpha$
indexed by $\alpha=1,\ldots,n,\bar n,\ldots,\bar1$. Three groups of indexes,
$\{1,\bar1\}$, $\{2,\bar2\}$ and $\{3,\bar3,\ldots,n,\bar n\}$, 
play separate roles in the product. 
An $R$ matrix $R_{i,j}$ acts on two components $V_i$ and $V_j$. 
We call the first component $V_i$ the auxiliary space and the second
component $V_j$ the quantum space. For any $R$ matrix contained in $A_\alpha$,
the auxiliary space is indexed by $1$ or $\bar1$, and the quantum space
by the other two groups. The second and the third groups are distinct
when the residue at $\lambda_{1,2}=0$ is calculated and the vector
$s^{(1)}_{\g2}s^{(1)}_{\og\overline{2}}h_{n-2}(\lambda_3,\ldots,\lambda_n)$ is created.


Our goal is to rewrite the product of $R$ matrices by means of the transfer
matrix 
$t^{(k)}(\lambda)=\tr_{V^{(k)}}\pi^{(k)}\(T^{[1]}(\lambda)\)\,
\in{\rm End}(W^{[1]})$, 
where $T^{[1]}(\lambda)$ is given by \eqref{eq:Tj}, 
and $W^{[1]}$ by \eqref{eq:Wj}. 
Let us compare the product of transfer matrices
$t^{(1)}(\lambda_1-k+1)\cdots t^{(1)}(\lambda_1)$ 
and the product
$A_{\bar1}(\lambda_1-k,\ldots)A_1(\lambda_1-k+1,\ldots)\cdots
A_\gamma(\lambda_1-1,\ldots)$. 
In the former, the 
matrix product is taken only on the quantum spaces indexed by
$2,\ldots,n,\bar n\ldots,\bar2$; 
in the latter not only on the quantum spaces
but also the auxiliary spaces indexed by $1$ and $\bar1$.

In a graphical representation of the product, we draw vertical
lines for quantum spaces and horizontal lines for auxiliary spaces. 
In the former product we have $k$ horizontal lines, and they form closed circles
reflecting the trace; in the latter we have only two horizontal lines
corresponding to $1$ and $\bar1$. They form spirals. 
We can rewrite 
the latter as a trace by introducing $k$ auxiliary spaces indexed by
$\alpha_1,\ldots,\alpha_k$. We do the following procedure.
We cut the horizontal lines
in front of the vertical line corresponding to $\bar3$. We obtain
$k$ separate horizontal lines. We rename these lines by
$\alpha_1,\ldots,\alpha_k$. Namely, we replace the spaces $V_1$ and $V_{\bar1}$
on these lines by $V_{\alpha_1},\ldots,V_{\alpha_k}$.
We recover the original product by taking the traces on these
new auxiliary spaces.

After some manipulation using the Yang-Baxter relation,
the whole expression comes in two parts: the fused monodromy operator
\[
\pi^{(k)}_{(\alpha_1,\cdots\alpha_k)}
\(T^{[1,2]}(\lambda_2-(k+1)/2)\)
\]
and the rest.
The latter acts on the vector 
$s^{(1)}_{\g2}s^{(1)}_{\og\overline{2}}$. 
This action can be rewritten as 
\[
\pi^{(k)}_{(\alpha_1,\cdots\alpha_k)}
\(L_2(-\frac{k+1}2)
L_{\bar2}(-\frac{k+3}2)\)s^{(2)}_{(1\bar1),(2\bar2)}.
\]
Combining these two expressions inside the trace
$\tr_{\alpha_1,\ldots,\alpha_k}$ we obtain (\ref{eq:1j}) for $j=2$
and $k\leq-3$. The calculation of the residues at the rest of the poles
can be done by using symmetries.

Let us set 
\bea
&&t_\alpha(\lambda_2)=
r_{\alpha\overline{3}}(\lambda_{23}-1)\cdots r_{\alpha\overline{n}}(\lambda_{2n}-1)
r_{\alpha n}(\lambda_{2n})\cdots r_{\alpha3}(\lambda_{23}),
\label{eq:talpha}
\ena
and recall that 
\be
&&
T^{[1,2]}(\la)=
L_{\bar{3}}(\la-\la_3-1)\cdots L_{\bar{n}}(\la-\la_n-1)
L_{n}(\la-\la_n)\cdots L_{3}(\la-\la_3), 
\\
&&T^{[1]}(\la)=L_{\bar{2}}(\la-\la_2-1)T^{[1,2]}(\la)L_{2}(\la-\la_2).
\en
By using (\ref{L4}), we obtain
\begin{eqnarray}\label{L5}
&&t_{\alpha_k}(\lambda_2-k)\cdots t_{\alpha_1}(\lambda_2-1)
\P^+_{\alpha_1,\cdots,\alpha_k}\\
&&=\frac{\prod_{j=3}^n\prod_{l=1}^k\{(\lambda_{2j}-l)^2-1\}}
{\prod_{j=3}^n\lambda_{2j}(\lambda_{2j}-k-1)}
\pi^{(k)}_{(\alpha_1,\cdots,\alpha_k)}
\left(T^{[1,2]}(\lambda_2-\frac{k+1}2)\right).
\nn
\end{eqnarray}
Here $\cP^+_{\al_1,\cdots,\al_k}$ stands for the projector
onto the completely symmetric tensors. 

Now we start the calculation.
We rewrite $A_\alpha$ given by (\ref{FUNA}):
\begin{equation}\label{E1}
A_\alpha(\lambda_1,\lambda_2,\ldots,\lambda_n)=
\frac{-1}{\prod_{j=2}^n(\lambda_{1j}^2-1)}
r_{\alpha\overline{2}}(\lambda_{12}-1)t_\alpha(\lambda_1)
r_{\alpha2}(\lambda_{12}),
\end{equation}
In this form it is easy to see
\begin{equation}
\res_{\lambda_{12}=-1}A_\alpha(\lambda_1,\ldots,\lambda_n)
=\frac{-1}{\prod_{j=3}^n\lambda_{2j}(\lambda_{2j}-2)}
r_{\alpha\overline{2}}(-2)t_\alpha(\lambda_2-1)\cP^-_{\al 2}.\label{RESA}
\end{equation}
Using \eqref{eq:rqKZ}, (\ref{RESA}) and (\ref{eq:n_to_n-2}) we obtain
\be
&&\res_{\lambda_{12}=-2}h_n(\lambda_1,\lambda_2,\ldots,\lambda_n)
_{12\cdots n\overline{n}\cdots\overline{2}\overline{1}}\\
&&=\frac{1}{2\prod_{j=3}^n\la_{2,j}(\la_{2,j}-2)}
r_{\overline{1}\overline{2}}(-2)t_{\overline{1}}(\lambda_2-1)s^{(1)}_{\overline{1}2}
s^{(1)}_{1\overline{2}}h_{n-2}(\lambda_3,\ldots,\lambda_n)
_{3\cdots n\overline{n}\cdots\overline{3}}.
\en

\medskip
Now we consider the residue at $\lambda_{12}=-k-1$
where $k\geq2$. We set
\begin{equation}
\g=
\begin{cases}
1&\hbox{ if $k$ is even;}\\
\overline{1}&\hbox{ if $k$ is odd.}
\end{cases}
\end{equation}
The only poles in (\ref{E1}) are $\lambda_{1j}=\pm1$.
By using \eqref{eq:rqKZ} repeatedly we obtain
\begin{eqnarray}
&&
\res_{\lambda_{12}=-k-1}h_n(\lambda_1,\lambda_2,\ldots,\lambda_n)
_{12\cdots n \overline{n}\cdots\overline{2}\overline{1}}
=X_kh_{n-2}(\lambda_3,\ldots,\lambda_n)_{3\cdots n\overline{n}\cdots\overline{3}},\label{E3}
\end{eqnarray}
for $k\geq1$, where
\begin{eqnarray}
&&X_k=\frac{1}{2\prod_{j=3}^n\la_{2,j}(\la_{2,j}-2)}
A_{\overline{1}}(\lambda_2-k,\lambda_2,\ldots,\lambda_n)
A_{1}(\lambda_2-k+1,\lambda_2,\ldots,\lambda_n)\label{Xk}\\
&&\times\cdots\times A_{\og}(\lambda_2-2,\lambda_2,\ldots,\lambda_n)
r_{\g\overline{2}}(-2)t_{\g}(\lambda_2-1)s^{(1)}_{\g2}s^{(1)}_{\og\overline{2}}.
\nn
\end{eqnarray}
If $k\geq2$, using (\ref{E1}) we can rewrite (\ref{Xk}) as
\begin{equation}\label{BULLET}
X_k=
\frac{(-1)^{k+1}}{(k-1)!(k+1)!\prod_{l=1}^k\prod_{j=3}^n((\la_{2,j}-l)^2-1)}
r_{\overline{1}\overline{2}}(-k-1)r_{1\overline{2}}(-k)\bullet Y_k
\end{equation}
where
\begin{eqnarray}
&&Y_k=
t_{\overline{1}}(\lambda_2-k)r_{\overline{1}2}(-k)r_{\overline{1}\overline{2}}(-k+1)
t_1(\lambda_2-k+1)r_{12}(-k+1)r_{1\overline{2}}(-k+2)
\label{E4}\\
&&\times\cdots\times t_\g(\lambda_2-3)r_{\g2}(-3)r_{\g\overline{2}}(-2)
t_{\og}(\lambda_2-2)r_{\og2}(-2)t_{\g}(\lambda_2-1)
s^{(1)}_{\g2}s^{(1)}_{\og\overline{2}}.
\nn
\end{eqnarray}

We use the following simple identity, which follows from \eqref{SINGL}.
\begin{equation}\label{L2}
r_{12}(-1)r_{13}(-2)r_{23}(-1)=0.
\end{equation}
\begin{lem}
We can insert $\P^+_{1\overline{1}}$ at the position $\bullet$ in \eqref{BULLET}. 
\end{lem}
\begin{proof}
Since $\P^+_{1\bar1}+\P^-_{1\bar1}={\rm id}_{1\bar1}$ and
$-2\P^-_{1\bar1}=r_{1\bar1}(-1)$, it is enough to show
that if we insert $r_{1\bar1}(-1)$ at the position $\bullet$ then
(\ref{BULLET}) vanishes. Note that $s^{(1)}=-\frac12r(-1)s^{(1)}$.
If $\gamma=1$, by using the Yang-Baxter relation, one can find
$r_{1\bar1}(-1)r_{\bar12}(-2)r_{12}(-1)$ therein; if $\gamma=\bar1$,
$r_{1\bar1}(-1)r_{1\bar2}(-2)r_{\bar12}(-1)$ instead.
\end{proof}
Then, by using (\ref{L6}) we have
\begin{equation}
X_k=\frac{(-1)^k k}{(k-1)!(k+1)!\prod_{l=1}^k\prod_{j=3}^n((\la_{2,j}-l)^2-1)}
r^{(2,1)}_{(1\overline{1}),\overline{2}}(-k-\frac12)Y_k.
\end{equation}

Now we use the identity
\begin{equation}
t_1(\la-j)=\tr_{V_\al}\left(t_\al(\la-j)P_{\al,1}\right).
\end{equation}
We can rewrite (\ref{E4}) as
\begin{eqnarray}
&&Y_k=\tr_{\alpha_1\cdots\alpha_k}\Bigl(
t_{\alpha_k}(\lambda_2-k)\cdots t_{\alpha_1}(\lambda_2-1)\bullet
\label{E4BIS}
\\
&&
\times r_{\alpha_k2}(-k)\cdots r_{\alpha_32}(-3)
r_{\alpha_k\overline{2}}(-k+1)\cdots r_{\alpha_3\overline{2}}(-2)\nn\\
&&\times r_{\alpha_22}(-2)P_{\alpha_k\overline{1}}P_{\alpha_{k-1}1}\cdots
P_{\alpha_2\og}P_{\alpha_1\g}\bullet\Bigr)
s^{(1)}_{\g2}s^{(1)}_{\og\overline{2}},\nn
\end{eqnarray}
where $\tr_{\alpha_1,\cdots,\alpha_k}$ stands
for the trace over $V_{\alpha_1}\otimes\cdots\otimes
V_{\alpha_k}$.
\begin{lem}\label{L3}
We can insert $\P^+_{\alpha_k\cdots\alpha_1}$ at the position
$\bullet$ in \eqref{E4BIS} for both places.
\end{lem}
\begin{proof}
We prove the assertion for the first $\bullet$. Then, the assertion for
the second $\bullet$ follows from the cyclicity of the trace.

Define an element of ${\rm End}(W)$,
\begin{equation}
\tilde Y_k={\rm tr}_{\alpha_k,\ldots,\alpha_1}
\left((t_k)_{\alpha_k,\ldots,\alpha_1}r_{\alpha_k,\ldots,\alpha_1;2,\bar2}
P_{\alpha_k\bar1}P_{\alpha_{k-1}1}\cdots
P_{\alpha_2\bar\gamma}P_{\alpha_1\gamma}\right)
\end{equation}
where
\begin{eqnarray}
&&(t_k)_{\alpha_k,\ldots,\alpha_1}
=t_{\alpha_k}(\lambda_2-k)\cdots t_{\alpha_1}(\lambda_2-1),\label{NONSYMT}\\
&&r_{\alpha_k,\ldots,\alpha_1;2,\bar2}
=r_{\alpha_k2}(-k)\cdots r_{\alpha_12}(-1)
r_{\alpha_k\bar2}(-k+1)\cdots r_{\alpha_2\bar2}(-1).
\end{eqnarray}

Let $t^{\rm sym}_k$ be the symmetrization of $t_k$,
\begin{equation}
t^{\rm sym}_k(v_{\tau_k}\otimes\cdots\otimes v_{\tau_1})
=\frac1{k!}\sum_{\sigma\in\mathfrak{S}_k}
t(\lambda_2-k)^{\tau'_k}_{\tau_{\sigma(k)}}\cdots
t(\lambda_2-1)^{\tau'_1}_{\tau_{\sigma(1)}}
v_{\tau'_k}\otimes\cdots\otimes v_{\tau'_1}.
\end{equation}
By the definition the matrix element
$\left(t^{\rm sym}_k\right)^{\tau'_k\cdots\tau'_1}_{\tau_k\cdots\tau_1}$
depends on $\tau_1,\ldots,\tau_k$ only by $\tau_k+\cdots+\tau_1=i$. We write it
$\left(t^{\rm sym}_k\right)^{\tau'_k\cdots\tau'_1}_i$.
We have the following symmetry of $t^{\rm sym}_k$.
\begin{equation}\label{FUSSYM}
\left(t^{\rm sym}_k\right)^{\tau'_k\cdots\tau'_1}_i
=\left(t^{\rm sym}_k\right)^{\tau'_{\sigma(k)}\cdots\tau'_{\sigma(1)}}_i.
\end{equation}

Let $\tilde Y'_k$ denote the matrix $\tilde Y_k$ in which $t_k$ is replaced
with $t^{\rm sym}_k$. We are to show that $\tilde Y_k=\tilde Y'_k$. 
We use the following equivalent definition of $\tilde Y_k$ (and a similar one for $\tilde Y'_k$).
\begin{eqnarray}
\tilde Y_k(v_{\varepsilon_1}\otimes v_{\varepsilon_{\bar1}})&=&
\sum_{{\varepsilon'_1},\varepsilon'_{\bar1}}
\sum_{\buildrel{\tau_1,\ldots,\tau_k}\over{\tau'_1,\ldots,\tau'_{k-2}}}
\left((t_k)_{\alpha_k,\ldots,\alpha_1}\right)
^{\varepsilon'_{\bar1}\varepsilon'_1\tau'_{k-2}\cdots\tau'_1}
_{\tau_k\tau_{k-1}\tau_{k-2}\cdots\tau_1}\\
&\times&\left(r_{\alpha_k,\ldots,\alpha_1;2,\bar2}\right)
^{\tau_k\tau_{k-1}\tau_{k-2}\cdots\tau_1}
_{\tau'_{k-2}\cdots\tau'_1\varepsilon_{\bar\gamma}\varepsilon_\gamma}
(v_{\varepsilon'_1}\otimes v_{\varepsilon'_{\bar1}}).\nonumber
\end{eqnarray}

We will show the following equality for each $i$.
\begin{eqnarray}\label{TOSHOW4.8}
&&\sum_{\buildrel{\buildrel{\tau_1,\ldots,\tau_k}\over{\tau_1+\cdots+\tau_k=i}}
\over{\tau'_1,\ldots,\tau'_{k-2}}}
\left((t_k)_{\alpha_k,\ldots,\alpha_1}\right)
^{\varepsilon'_{\bar1}\varepsilon'_1\tau'_{k-2}\cdots\tau'_1}
_{\tau_k\tau_{k-1}\tau_{k-2}\cdots\tau_1}
\left(r_{\alpha_k,\ldots,\alpha_1;2,\bar2}\right)
^{\tau_k\tau_{k-1}\tau_{k-2}\cdots\tau_1}
_{\tau'_{k-2}\cdots\tau'_1\varepsilon_{\bar\gamma}\varepsilon_\gamma}\\
&&=
\sum_{\buildrel{\tau_1,\ldots,\tau_k}\over{\tau_1+\cdots+\tau_k=i}}
\left\{\sum_{\tau'_1,\ldots,\tau'_{k-2}}
(t^{\rm sym}_k)_i
^{\varepsilon'_{\bar1}\varepsilon'_1\tau'_{k-2}\cdots\tau'_1}
\left(r_{\alpha_k,\ldots,\alpha_1;2,\bar2}\right)
^{\tau_k\tau_{k-1}\tau_{k-2}\cdots\tau_1}
_{\tau'_{k-2}\cdots\tau'_1\varepsilon_{\bar\gamma}\varepsilon_\gamma}\right\}.
\nonumber
\end{eqnarray}
We prove the equality by using induction on $k$. The cases $k=2,3$
is immediately follows: 
By using the equalities
\begin{eqnarray}
&&r_{\alpha_2,\alpha_1}(-1)r_{\alpha_2,2}(-2)r_{\alpha_1,2}(-1)=0,\label{121}\\
&&r_{\alpha_3\alpha_2}(-1)r_{\alpha_32}(-3)r_{\alpha_22}(-2)
r_{\alpha_3\bar2}(-2)r_{\alpha_2\bar2}(-1)=0,\label{13221}
\end{eqnarray}
we see that $\left(r_{\alpha_k,\ldots,\alpha_1;2,\bar2}\right)
^{\tau_k\tau_{k-1}\tau_{k-2}\cdots\tau_1}
_{\tau'_{k-2}\cdots\tau'_1\varepsilon_{\bar\gamma}\varepsilon_\gamma}$
is totally symmetric in $\tau_1,\ldots\tau_k$.
For $k\geq4$ the last expression is not totally symmetric.

The equality is clear if $i=k$ or $i=-k$ because there is only one term
for the summation of $\tau_1,\ldots,\tau_k$,
and $t_k$ and $t^{\rm sym}_k$ are the same in this sector.
Now, assume that $i\not=\pm k$.

Using the Yang-Baxter equation, and the equalities \eqref{121}, \eqref{13221}
we see that the summand $\{\cdots\}$ in the right hand side
of \eqref{TOSHOW4.8} is independent of $\tau_1,\tau_2,\cdots,\tau_k$.
Since $i\not=\pm k$, we can choose it as $\tau_k=+$ and $\tau_{k-1}=-$.
Then, the right hand side becomes
\begin{eqnarray}
&&\frac{4k(k-1)}{(k-i)(k+i)}
\sum_{\buildrel{\tau_1,\ldots,\tau_{k-2}}\over{\tau_1+\cdots+\tau_{k-2}=i}}
\Bigl\{\sum_{\tau'_1,\ldots,\tau'_{k-2}}
\Bigl(\frac{k+i}{2k}t(\lambda_2-k)^{\varepsilon'_{\bar1}}_+
(t^{\rm sym}_{k-1})_{i-1}^{\varepsilon'_1\tau'_{k-2}\cdots\tau'_1}\\
&&     +\frac{k-i}{2k}t(\lambda_2-k)^{\varepsilon'_{\bar1}}_-
(t^{\rm sym}_{k-1})_{i+1}^{\varepsilon'_1\tau'_{k-2}\cdots\tau'_1}\Bigr)
\left(r_{\alpha_k,\ldots,\alpha_1;2,\bar2}\right)
^{+-\tau_{k-2}\cdots\tau_1}
_{\tau'_{k-2}\cdots\tau'_1\varepsilon_{\bar\gamma}\varepsilon_\gamma}\Bigr\}.
\nonumber
\end{eqnarray}
Consider the first half,
\begin{equation}\label{FHALF}
\frac{2(k-1)}{k-i}\hskip-15pt
\sum_{\buildrel{\tau_1,\ldots,\tau_k}\over{\tau_1+\cdots+\tau_{k-2}=i}}
\hskip-5pt
\Bigl\{\sum_{\tau'_1,\ldots,\tau'_{k-2}}
t(\lambda_2-k)^{\varepsilon'_{\bar1}}_+
(t^{\rm sym}_{k-1})_{i-1}^{\varepsilon'_1\tau'_{k-2}\cdots\tau'_1}
\left(r_{\alpha_k,\ldots,\alpha_1;2,\bar2}\right)
^{+-\tau_{k-2}\cdots\tau_1}
_{\tau'_{k-2}\cdots\tau'_1\varepsilon_{\bar\gamma}\varepsilon_\gamma}\Bigr\}.
\end{equation}
By the same argument as before, we see that the summand $\{\cdots\}$
is invariant for the permutation of the indices $-,\tau_{k-2},\cdots,\tau_1$.
Therefore, we can rewrite \eqref{FHALF} as
\begin{equation}
\sum_{\buildrel{\tau_1,\ldots,\tau_{k-1}}\over{\tau_1+\cdots+\tau_{k-1}=i-1}}
\Bigl\{\sum_{\tau'_1,\ldots,\tau'_{k-2}}
t(\lambda_2-k)^{\varepsilon'_{\bar1}}_+
(t^{\rm sym}_{k-1})_{i-1}^{\varepsilon'_1\tau'_{k-2}\cdots\tau'_1}
\left(r_{\alpha_k,\ldots,\alpha_1;2,\bar2}\right)
^{+\tau_{k-1}\tau_{k-2}\cdots\tau_1}
_{\tau'_{k-2}\cdots\tau'_1\varepsilon_{\bar\gamma}\varepsilon_\gamma}\Bigr\}.
\end{equation}
Now, by induction hypothesis, this is equal to
\begin{equation}\label{FHALF2}
\sum_{\buildrel{\tau_1,\ldots,\tau_{k-1}}\over{\tau_1+\cdots+\tau_{k-1}=i-1}}
\Bigl\{\sum_{\tau'_1,\ldots,\tau'_{k-2}}
(t_k)_{+\tau_{k-1}\cdots\tau_1}
^{\varepsilon'_{\bar1}\varepsilon'_1\tau'_{k-2}\cdots\tau'_1}
\left(r_{\alpha_k,\ldots,\alpha_1;2,\bar2}\right)
^{+\tau_{k-1}\tau_{k-2}\cdots\tau_1}
_{\tau'_{k-2}\cdots\tau'_1\varepsilon_{\bar\gamma}\varepsilon_\gamma}\Bigr\}.
\end{equation}
In order to rewrite the second half in a similar manner, we should first
rewrite the indices $+-$ in 
$\left(r_{\alpha_k,\ldots,\alpha_1;2,\bar2}\right)
^{+-\tau_{k-2}\cdots\tau_1}
_{\tau'_{k-2}\cdots\tau'_1\varepsilon_{\bar\gamma}\varepsilon_\gamma}$
to $-+$. This is possible by the same argument again.
After that, the argument is similar, and we obtain a similar expression
to \eqref{FHALF2}. Adding these two expressions,
we obtain the left hand side of \eqref{TOSHOW4.8}.
\end{proof}
Using (\ref{L5}), we obtain
\begin{eqnarray}
&&
X_k=d_kr^{(2,1)}_{(1\overline{1}),\overline{2}}(-k-\frac12)
\tr_{\alpha_1\cdots\alpha_k}\Bigl\{
\pi^{(k)}_{(\alpha_1\cdots\alpha_k)}
\left(T^{[1,2]}(\lambda_2-\frac{k+1}2)\right)
\label{E5}\\
&&
\times r_{\alpha_k2}(-k)\cdots r_{\alpha_32}(-3)
r_{\alpha_k\overline{2}}(-k+1)\cdots r_{\alpha_3\overline{2}}(-2)\bullet
\nn\\
&&\times P_{\alpha_2\og}P_{\alpha_1\g}
\P^+_{\alpha_k\cdots\alpha_1}\Bigr\}
r_{\og2}(-2)s^{(1)}_{\g2}s^{(1)}_{\og\overline{2}},
\nn
\end{eqnarray}
where
\begin{equation}
d_k=\frac{(-1)^kk}
{(k-1)!(k+1)!\prod_{j=3}^n\lambda_{2j}
(\lambda_{2j}-k-1)}.
\end{equation}
We insert $\P^+_{\alpha_k\cdots\alpha_3}$ and $\P^+_{2\overline{2}}$
at the position $\bullet$ in (\ref{E5}), and using (\ref{L4}) and (\ref{L6})
obtain
\begin{eqnarray}
&&X_k=\frac{(k-2)!(k-1)!d_k}2\,r^{(2,1)}_{(1\overline{1}),\overline{2}}(-k-\frac12)
\\
&&\times\tr_{\alpha_1\cdots\alpha_k}\Bigl\{
\pi^{(k)}_{(\alpha_1\cdots\alpha_k)}
\Bigl(T^{[1,2]}\bigl(\lambda_2-\frac{k+1}2\bigr)\Bigr)
r^{(k-2,2)}_{(\alpha_3\cdots\alpha_k),(2\overline{2})}(-\frac{k+2}2)
\nn\\
&&\times P_{\alpha_2\og}P_{\alpha_1\g}
\P^+_{\alpha_1\cdots\alpha_k}\Bigr\}
r_{\og2}(-2)s^{(1)}_{\g2}s^{(1)}_{\og\overline{2}}.
\nn
\end{eqnarray}

Now we rewrite the last part. First we use
\begin{equation}\label{L7}
r_{\og2}(-2)s^{(1)}_{\g2}s^{(1)}_{\og\overline{2}}
=-2s^{(2)}_{(1\overline{1}),(2\overline{2})}.
\end{equation}
Then, we use
\begin{eqnarray}\label{L8}
P_{\alpha_2\og}P_{\alpha_1\g}\P^+_{\alpha_1,\alpha_2}s^{(2)}_{(1\overline{1}),(2\overline{2})}
&=&\frac12r^{(2,2)}_{(\alpha_1\alpha_2),(1\overline{1})}(0)
s^{(2)}_{(1\overline{1}),(2\overline{2})}
\\
&=&\frac12r^{(2,2)}_{(\alpha_1\alpha_2),(2\overline{2})}(-1)s^{(2)}_{(1\overline{1}),(2\overline{2})}.
\nn
\end{eqnarray}
where we used the crossing symmetry (\ref{eq:crossing}).

Finally, applying (\ref{L9}) to
$r^{(k-2,2)}_{(\alpha_3\cdots\alpha_k),(2\overline{2})}(-\frac{k+2}2)$
and 
$r^{(2,2)}_{(\alpha_1\alpha_2),(2\overline{2})}(-1)$,
we obtain
\begin{eqnarray}\label{SUMMARY1}
\\&&X_k=\frac {(-1)^{k+1}}
{(k^2-1)\prod_{j=3}^n\lambda_{2j}(\lambda_{2j}-k-1)}\nn\\
&&\times
\tr_{\alpha_1\cdots\alpha_k}\left\{\pi^{(k)}_{(\alpha_1\cdots\alpha_k)}
\left(T^{[1,2]}(\lambda_2-\frac{k+1}2)
r^{(2,1)}_{(1,\bar{1}),\bar{2}}(-k-\frac{1}{2})
L^{(2)}_{(2,\bar{2})}\(-\frac{k}{2}\)
\right)\right\}s^{(2)}_{(1,\bar{1}),(2,\bar{2})}.\nn
\end{eqnarray}
This expression can be further simplified using 
Lemma \ref{LemL}. 
Using the cyclicity of trace we come to the final result:
\begin{eqnarray}
&&X_k=\frac {(-1)^{k+1}}
{(k+1)\prod_{j=3}^n\lambda_{2j}(\lambda_{2j}-k-1)}\label{SUMMARY2}\\
&&\times
\tr_{\alpha_1\cdots\alpha_k}\left\{\pi^{(k)}_{(\alpha_1\cdots\alpha_k)}
\left(T^{[1]}(\lambda_2-\frac{k+1}2)
\right)\right\}s^{(2)}_{(1,\bar{1}),(2,\bar{2})}.\nn
\end{eqnarray}

Let us summarize the result by using
the function $\omega$ and $X^{[1,2]}$.
\begin{lem}\label{lem:diff_eq}
The following difference equation holds for $X^{[1,2]}$:
\bea
&&X^{[1,2]}(\la_1-1,\la_2,\cdots,\la_n) 
\label{eq:diff}
\\
&&=-\frac{(\lambda_{12}-1)(\lambda_{12}+1)}{\lambda_{12}(\lambda_{12}-2)}
A_{\bar{1}}(\la_1,\cdots,\la_n)P_{1\bar{1}}
X^{[1,2]}(\la_1,\la_2,\cdots,\la_n). \nn 
\ena
\end{lem}
\begin{proof}
Eq. \eqref{SUMMARY2} shows that 
\bea
X_k=(-1)^kk(k+2)X^{[1,2]}(\lambda_2-k-1,\lambda_2,\ldots,\lambda_n)
\label{TOSHOW}
\ena
for all integers $k\geq 2$.  
On the other hand, from the definition of $X_k$ we have
\bea
X_{k}=
A_{\overline{1}}(\lambda_{2}-k, \lambda_{2}, \ldots, \lambda_{n})P_{1\overline{1}}X_{k-1}.\label{eq:Xkrec}
\ena
Therefore (\ref{eq:diff}) is valid if $\lambda_{1,2}=-k$ 
for all integers $k\ge 2$. 
Since both sides of (\ref{eq:diff}) are rational, 
it is valid identically. 
\end{proof}

\begin{prop}\label{prop:residue}
For any positive integer $k\geq1$ we have
\bea
&&
\res_{\la_1=\la_2-k-1}h_n(\la_1,\cdots,\la_n)_{1,\cdots, n,\bar n,\cdots,\bar1}
\label{eq:residue}\\
&&=\res_{\la_1=\la_2-k-1}\{\omega(\la_{1,2})
X^{[1,2]}(\la_1,\cdots,\la_n)\}
h_{n-2}(\la_3,\cdots,\la_n)_{3,\cdots,n,\bar{n},\cdots,\bar{3}}.
\nn
\ena
\end{prop}
\begin{proof}
Note that $X^{[1,2]}(\lambda_1,\lambda_2,\ldots,\lambda_n)$
has no pole at $\lambda_1=\lambda_2-k-1$ with $k\ge 1$.   
Since \eqref{eq:Xkrec} uniquely determines $X_{k-1}$ from $X_k$ for 
$k\ge 2$, the difference equation \eqref{eq:diff}
implies that 
\eqref{TOSHOW} is valid also for $k=1$.
Eq. \eqref{eq:residue} is then 
an immediate consequence of 
\eqref{TOSHOW}
and
\begin{equation}
{\rm res}_{\lambda=-k-1}\omega(\lambda)=(-1)^kk(k+2).
\end{equation}
\end{proof}

It remains to find residues at poles $\la _1=\la_2+k+1$ 
for $k\ge 1$. The result is formulated in the
\begin{prop}\label{prop:residue1}
For any positive integer $k\geq1$ we have
\bea
&&
\res_{\la_1=\la_2+k+1}h_n(\la_1,\cdots,\la_n)_{1,\cdots ,n,\bar n,\cdots,\bar1}
\label{eq:residue2}\\
&&=\res_{\la_1=\la_2+k+1}\{\omega(\la_{1,2})
X^{[1,2]}(\la_1,\cdots,\la_n)\}
h_{n-2}(\la_3,\cdots,\la_n)_{3,\cdots,n,\bar{n},\cdots,\bar{3}}.
\nn
\ena
\end{prop}
\begin{proof}
Noting that $\omega(\la)$ is even, we have
\begin{align}
&\res_{\la_1=\la_2+k+1}h_n(\la_1,\la _2\cdots,\la_n)_{1,2\cdots n\bar n\cdots\bar2,\bar1}\nn\\&=
\res_{\la_1=\la_2+k+1}R_{2,1}(\la _{21})R_{\bar1,\bar2}(\la_{12})
h_n(\la_2,\la_1,\cdots,\la_n)_{2,1\cdots n\bar n\cdots\bar1,\bar2}\nn\\&
=R_{2,1}(-k-1)R_{\bar1,\bar2}(k+1)
\res_{\la_1=\la_2+k+1}\omega(\la_{1,2})
\nn\\&\times
X^{[1,2]}(\la_2,\la _2+k+1, \cdots,\la_n)_{2,1\cdots n\bar n\cdots\bar1,\bar2}
h_{n-2}(\la_3,\cdots,\la_n)_{3,\cdots,n,\bar{n},\cdots,\bar{3}}\nn
\end{align}
Hence the assertion follows from the symmetry \eqref{SymX}.
\end{proof}

\begin{cor}\label{prop:residue2}
For any $k\in \Z\backslash\{0,\pm 1\}$ and $2\le j\le n$ we have
\bea
&&\label{eq:residue3} \\ 
&&
\res_{\la_1=\la_j-k}h_n(\la_1,\cdots,\la_n)_{1,\cdots ,n,\bar n,\cdots,\bar1}
\nn\\
&&=\res_{\la_1=\la_j-k}\{\omega(\la_{1,j})
X^{[1,j]}(\la_1,\cdots,\la_n)\}
h_{n-2}(\la_2,\cdots,\widehat{\la_j},\cdots,\la_n)
_{2,\cdots,\hat{j},\cdots,n,\bar{n},\cdots,\hat{\bar{j}},\cdots,\bar{2}}.
\nn
\ena
\end{cor}
\begin{proof}
This is an immediate consequence of 
Proposition \ref{prop:residue}--\ref{prop:residue1},
the symmetry \eqref{eq:Rsymm} and the relation \eqref{eq:X2Xj}.  
\end{proof}


\subsection{Asymptotics}\label{subsec:4.4}
In this subsection we finish the proof of Theorem \ref{thm:recursion}.
Let $\Phi_L(\la_1)=h_n(\la_1,\cdots,\la_n)$, 
$\Phi_R(\la_1)$ the right hand side of \eqref{eq:Recursion},
and set $\Phi(\la_1)=\Phi_L(\la_1)-\Phi_R(\la_1)$.   
We are going to show that $\Phi(\la_1)=0$. 

By Corollary \ref{prop:residue2}, we have 
\be
\res_{\la_1=\la_j-k}\Phi_L(\la_1)
=\res_{\la_1=\la_j-k}\Phi_R(\la_1)
\en
for all $k\in\Z\backslash\{0,\pm 1\}$ and $j=2,\cdots,n$. 
By \eqref{eq:meromorphy} and the definition of 
$Z^{[1,j]}(\la_1,\cdots,\la_n)$, there are no other poles in $\la_1$.  
Hence $\Phi(\la_1)$ is an entire function. 

Consider the asymptotic behavior  as $\la_1\to\infty$.  
By \eqref{eq:reg_at_infty} we know that 
\be
\lim_{\la_1\to\infty\atop \la_1\in S_\delta}
\Phi_L(\la_1)
=(-1)^{n-1}\frac{1}{2}s_{1,\bar{1}}h_{n-1}(\la_2,\cdots,\la_n)
\en
for any $0<\delta<\pi$, 
where $S_\delta=\{\la\in \C\mid \delta<|{\rm arg}\,\la_1|<\pi-\delta\}$.
On the other hand, the function $\omega(\la)$ satisfies 
\be
\lim_{\la_1\to\infty\atop \la_1\in S_\delta}\omega(\la)=0. 
\en
Since the coefficients of $\omega(\la_{1,j})$ are 
regular at $\la_1=\infty$ (Lemma \ref{lem:degree_of_X}), 
the terms in $\Phi_R(\la_1)$ except the last  
vanish in this limit. 
The last term of $\Phi_R(\la_1)$ is so chosen that 
\bea
\lim_{\la_1\to\infty\atop \la_1\in S_\delta}\Phi(\la_1)=0.
\label{eq:sector}
\ena
Let us show that the condition $\la_1\in S_\delta$ can be removed. 

\begin{lem}
There exist constants $M,c>0$ such that 
\be
|\Phi(\la_1)|\le M\,e^{c |\la_1|}\qquad (\la_1\in\C). 
\en
\end{lem}
\begin{proof}
Recall that $\Phi_L(\la_1)$ satisfies the difference equation of the form
$\Phi_L(\la_1+1)=B(\la_1)\Phi_L(\la_1)$, where
$B(\la_1)$ is a matrix of rational functions 
holomorphic and invertible at $\la_1=\infty$. 
Take a small neighborhood $U$ of the set of poles of $B(\la_1)^{\pm 1}$, 
and choose 
$K'>0$ such that 
$\sup_{\la_1\in \mathbb{P}^1\setminus U}|B(\la_1)^{\pm 1}|\le K'$. 
Choose $\la_1^0\in\C$ so that 
$\la_1^0+\Z+i\R\subset \mathbb{P}^1\setminus U$. 
Then for $n\ge 0$ we have 
\bea
|\Phi_L(\la_1^0+it+n)|&\le& |B(\la_1^0+it+n-1)|\cdots |B(\la_1^0+it)|
|\Phi_L(\la_1^0+it)|
\label{eq:L}\\
&\le& M' {K'}^n,
\nn
\ena
where $M'=\sup_{t\in\R}|\Phi_L(\la_1^0+it)|$. Clearly
a similar estimate holds also for $|\Phi_L(\la_1^0+it-n)|$. 

On the other hand, the function $\omega(\la)$ 
satisfies the difference equation
\be
\begin{pmatrix}
\omega(\la+1)\\ 1\\
\end{pmatrix}
=\begin{pmatrix}
-\frac{\la(\la+2)}{\la^2-1} & \frac{3}{2(\la^2-1)} \\
0 & 1 \\
\end{pmatrix}
\begin{pmatrix}
\omega(\la)\\ 1\\
\end{pmatrix}.
\en
Hence by the same argument as above, we obtain 
an estimate for $\omega(\la_{1,j})$, and hence for 
$\Phi_R(\la_1)$, of the type \eqref{eq:L}.

In summary, there exist $M,K>0$ such that 
\be
\sup_{t\in\R}|\Phi(\la_1^0+it+n)|\le M K^{|n|}\qquad (n\in\Z).
\en
The Lemma follows from this and the maximum principle. 
\end{proof}

Now fix $\delta<\pi/2$.  
By \eqref{eq:sector} there exists an $M'\ge M$ such that 
\bea
|\Phi(\la_1)|\le M' 
\qquad
(\la_1\in\C, \delta\le |\arg \la_1|\le \pi-\delta). 
\label{eq:estimate}
\ena
By the lemma above and the Phragmen-Lindel{\" o}f theorem, 
\eqref{eq:estimate} holds also for $|\arg\la_1|\le\delta$ or 
$\pi-\delta\le|\arg\la_1|\le \pi$. 
Therefore $\Phi(\la_1)$ is bounded in the full neighborhood of $\la_1=\infty$. 
Hence $\la_1=\infty$ is a regular point of $\Phi(\la_1)$, 
and we conclude that $\Phi(\la_1)=0$.  
This completes the proof of Theorem \ref{thm:recursion}.


\appendix
\section{Relation to the correlators of the XXZ model}\label{app:A}

We give the relation between two different gauges, the one used in the 
XXZ model in the massive regime \cite{JM}, and the one used
in the present paper, which gives the Hamiltonian (\ref{APPA1}).

In \cite{JM}, the following Hamiltonian was considered:
\be
H_{XXZ}=-\frac{1}{2}\sum_{j=1}^L
\left(\sigma_j^x\sigma_{j+1}^x
+\sigma_j^y\sigma_{j+1}^y-\frac{x+x^{-1}}{2}\sigma_j^z\sigma_{j+1}^z\right).
\en
Here $q=-x$, the XXX limit is $x\to 1$. 
The gauge transformation by $\cK=\prod_{j:{\rm even}}\sigma_j^z$ 
brings $H_{XXZ}$ to 
\be
\widetilde{H}_{XXZ}&=&\cK H_{XXZ} \cK^{-1}
\\
&=&\frac{1}{2}\sum_{j=1}^L
\left(\sigma_j^x\sigma_{j+1}^x
+\sigma_j^y\sigma_{j+1}^y+\frac{x+x^{-1}}{2}\sigma_j^z\sigma_{j+1}^z\right).
\en
The ground states and the correlators are related as
\be
&& 
\ket{\widetilde{\Omega}}=\cK\ket{\Omega},
\\
&&
\bra{\widetilde{\Omega}}
(E_{\e_1,\eb_1})_1\cdots (E_{\e_n,\eb_n})_n
\ket{\widetilde{\Omega}}
=\prod_{1\le j\le n\atop j:{\rm even}}\e_j\eb_j
\cdot 
\bra{\Omega}(E_{\e_1,\eb_1})_1\cdots (E_{\e_n,\eb_n})_n
\ket{\Omega}.
\en

\be
G_{2n}(\zeta_1,\cdots,\zeta_n,x\zeta_n,\cdots,x\zeta_1)
^{-\e_1,\cdots,-\e_n,\eb_n,\cdots,\eb_1}
=
\bra{\Omega}(E_{\e_1,\eb_1})_1\cdots (E_{\e_n,\eb_n})_n
\ket{\Omega}.
\en

\be
g_{2n}(\zeta_1,\cdots,\zeta_{2n})
=
(-1)^n\prod_{1\le j\le 2n\atop j:{\rm even}}\sigma^z_j
\cdot
G_{2n}(\zeta_1,\cdots,\zeta_{2n}).
\en
Therefore,
\be
&&g_{2n}(\zeta_1,\cdots,\zeta_n,x\zeta_n,\cdots,x\zeta_1)
^{-\e_1,\cdots,-\e_n,\eb_n,\cdots,\eb_1}
\\
&&=
(-1)^{[n/2]}\prod_{j=1}^n(-\eb_j)
\cdot
\bra{\widetilde{\Omega}}
(E_{\e_1,\eb_1})_1\cdots (E_{\e_n,\eb_n})_n
\ket{\widetilde{\Omega}}.
\en
The Hamiltonian $\widetilde{H}_{XXZ}$ and the ground state
$\ket{\widetilde{\Omega}}$ are related to those in Section \ref{sec:2}
by
\[
{\rm lim}_{x\rightarrow1}\widetilde{H}_{XXZ}=H_{XXX},\quad
{\rm lim}_{x\rightarrow1}\ket{\widetilde{\Omega}}=|{\rm vac}\rangle.
\]
Therefore, we have the relation (\ref{APPA2}).
\section{Analyticity and asymptotic properties of $h_{n}$}\label{app:B}

Here we give a proof for Proposition \ref{prop:3.2}. 
We start from the integral formula for the function  
$h_{n}(\lambda_{1}, \ldots , \lambda_{n})
^{\epsilon_{1},\ldots,\epsilon_{n},\bar{\epsilon_{n}},
\ldots,\bar{\epsilon_{1}}}$
given in \cite{JM}. 

Set 
\be 
A=\{j| \epsilon_{j}=+ \}, \quad 
B=\{j | \bar{\epsilon_{j}}=+ \}. 
\en 
We write $A=\{a_{1}, \ldots , a_{r}\}, 
B=\{b_{1}, \ldots , b_{s}\}$ 
with $a_{1}< \cdots <a_{r}, b_{1}< \cdots <b_{s}$. 
Note that $r+s=n$. 
For each $a \in A$ and $b \in B$ we prepare 
an integration variable $t_a$ and $t_{b}'$, respectively. 
Arrange the variables as 
\be 
(u_{1}, \ldots , u_{n})=(t_{a_{r}}, \ldots , t_{a_{1}}, t_{b_{1}}',
\ldots , t_{b_{s}}'). 
\en 
Then the formula is given by 
\begin{eqnarray}
&& 
h_{n}(\lambda_{1}, \ldots , \lambda_{n})
^{\epsilon_{1}, \ldots , \epsilon_{n}, \bar{\epsilon}_{n}, \ldots ,
\bar{\epsilon}_{1}} \label{eq:integral-formula}\\ 
&& 
{}=c_{A, B}^{(n)}
\prod_{a \in A}\int_{C^{+}}
\frac{dt_{a}}{t_{a}-\lambda_{a}}
\prod_{b \in B}\int_{C^{-}}
\frac{dt_{b}'}{t_{b}'-\lambda_{b}} \nn \\ 
&& \qquad 
{}\times 
\prod_{a \in A \atop j<a}\frac{t_{a}-\lambda_{j}-1}{t_{a}-\lambda_{j}}
\prod_{b \in B \atop
j<b}\frac{t_{b}'-\lambda_{j}+1}{t_{b}'-\lambda_{j}} \nn \\ 
&& \qquad 
{}\times 
\prod_{j<k}\left(
\frac{\sinh{\pi i (u_{j}-u_{k})}}{u_{j}-u_{k}-1}
\frac{\sinh{\pi i (\lambda_{j}-\lambda_{k})}}{\lambda_{j}-\lambda_{k}}
\right) 
\prod_{j, k}\frac{u_{j}-\lambda_{k}}{\sinh{\pi i(u_{j}-\lambda_{k})}}.
\nn  
\end{eqnarray}
Here $c_{A, B}^{(n)}$ is a constant which depends on 
$A, B$ and $n$. 
The integration contour $C^{+}$ is parallel to 
the imaginary axis for $|{\rm Im}t_{a}|\gg 0$, and 
separates the sequences of the poles of the integrand 
into the two sets  
$\lambda_{j}+\mathbb{Z}_{\le 0}$ 
and $\lambda_{j}+\mathbb{Z}_{>0}$. 
Similarly $C^{-}$ is the contour separating the poles 
into $\lambda_{j}+\mathbb{Z}_{<0}$ and 
$\lambda_{j}+\mathbb{Z}_{\ge 0}$. 

Now let us check the analyticity property \eqref{eq:meromorphy}. 
The singularity of the integral 
\eqref{eq:integral-formula} comes from 
the pinch of the integration contour by some poles 
of the integrand. 
Hence the function $h_{n}$ is meromorphic 
with at most poles at 
$\lambda_{i}-\lambda_{j} \in\mathbb{Z}\setminus\{0\}$. 
Let us prove that $h_{n}$ is analytic at 
$\lambda_{i}=\lambda_{j}\pm 1$. 
{}From \eqref{eq:Rsymm} it suffices to prove that 
$h_{n}$ is analytic at $\lambda_{1}=\lambda_{2}\pm 1$. 
The function $h_{n}$ satisfies \eqref{eq:rqKZ}, and 
$A_{\bar{1}}(\lambda_{1}, \lambda_{2}, \ldots )$ and 
$h_{n}(\lambda_{1}, \lambda_{2}, \ldots )$ are 
regular at $\lambda_{1}=\lambda_{2}$. 
Hence $h_{n}$ is analytic at $\lambda_{1}=\lambda_{2}-1$. 
Similarly the regularity at $\lambda_{1}=\lambda_{2}+1$ can be checked 
by using the equation 
\begin{eqnarray*}
&& 
h_{n}(\lambda_{1}+1, \lambda_{2}, \ldots , \lambda_{n})_{1, \ldots , n,
\bar{n}, \ldots , \bar{1}} \\ 
&=&
\frac{-1}{\prod_{j=2}^{n}(\lambda_{1j}^{2}-1)}
r_{12}(-\lambda_{12}-1) \cdots 
r_{1n}(-\lambda_{1n}-1)
r_{1\bar{n}}(-\lambda_{1n}) \cdots 
r_{1\bar{2}}(-\lambda_{12}) \\ 
&& {}\times 
h_{n}(\lambda_{1}, \ldots , \lambda_{n})_{\bar{1}, 2, \ldots , n,
\bar{n}, \ldots , \bar{2}, 1}, 
\end{eqnarray*}
which is derived from \eqref{eq:rqKZ}. 
It is easy to check that the poles at 
$\lambda_{i}-\lambda_{j} \in \mathbb{Z}\setminus \{0, \pm1\}$ 
are simple by using \eqref{eq:rqKZ} repeatedly. 

Next we show the asymptotic property \eqref{eq:reg_at_infty}. 
We consider the asymptotics 
in the angular domain 
\bea
\la_1\to\infty, \quad 
\delta < |\rm{arg}\lambda_{1}| < \pi -\delta,
\label{eq:ang-domain}
\ena 
where $0<\delta<\pi$. 
Setting $\lambda_{1}=\kappa+i\mu$ with $\kappa, \mu \in \mathbb{R}$ 
we have in this domain 
\be
|\mu|\le |\la_1|\le K|\mu| 
\en
for a positive constant $K$. 
We will consider only the case $\mu\to+\infty$ 
since the other case is similar.  

First we deform the contours $C^{\pm}$ suitably and assume that 
the contours $C^{\pm}, C^{\pm}+1$ and $C^{\pm}-1$ do not cross each other. 
Taking $\mu \gg 0$ we also assume that $|{\rm Im}\lambda_{j}| \le \mu/6$ 
for $j=2, \ldots , n$, and that $C^{\pm}$ contains the segment
\be 
\Gamma^{\pm}=\{\kappa\pm\epsilon+iy | \frac{2}{3}\mu \le y \le \frac{4}{3}\mu \} 
\en 
for small $\epsilon>0$. 

Denote by $P_{A, B}$ the rational part of the integrand, that is 
\begin{eqnarray*}
&& 
P_{A, B}(u_{1}, \ldots , u_{n}; \lambda_{1}, \ldots , \lambda_{n}) \\ 
&&{}=
\prod_{a \in A}\left(\frac{1}{t_{a}-\lambda_{a}}
\prod_{j<a}\frac{t_{a}-\lambda_{j}-1}{t_{a}-\lambda_{j}} \right)  
\prod_{b \in B}\left(\frac{1}{t_{b}'-\lambda_{b}} 
\prod_{j<b}\frac{t_{b}'-\lambda_{j}+1}{t_{b}'-\lambda_{j}}\right) \\ 
&& \quad {}\times 
\frac{\prod_{j, k}(u_{j}-\lambda_{k})}
{\prod_{j<k}(u_{j}-u_{k}-1)(\lambda_{j}-\lambda_{k})}. 
\end{eqnarray*}
We use  
\begin{eqnarray*}
\prod_{j<k}\sinh{\pi i(u_{j}-u_{k})}&=&
2^{-{n-2 \atopwithdelims() 2}}e^{\pi i(n-1)\lambda_{1}} \\ 
&{}\times& 
{\rm Skew}\left(
\prod_{j=1}^{n-1}e^{\pi i(n-2j)u_{j}}\sinh{\pi i(u_{j}-\lambda_{1})}
\cdot e^{\pi i(-n+1)u_{n}} \right),  
\end{eqnarray*}
where ${\rm Skew}$ is the skew-symmetrization 
with respect to $u_{1}, \ldots , u_{n}$. 
Then we find 
\begin{eqnarray}
&& 
h_{n}(\lambda_{1}, \ldots , \lambda_{n})
^{\epsilon_{1}, \ldots , \epsilon_{n}, \bar{\epsilon}_{n}, \ldots ,
\bar{\epsilon}_{1}} \label{eq:asymp-decomp} \\ 
&& {}=
2^{-{n-2 \atopwithdelims() 2}}c_{A, B}^{(n)} \, 
e^{\pi i(n-1)\lambda_{1}}
\prod_{j<k}\sinh{\pi i(\lambda_{j}-\lambda_{k})}  \nn \\
&& \quad {}\times \sum_{\sigma \in \mathfrak{S}_{n}}({\rm sgn}\sigma)
\prod_{j=1}^{n}\int_{C_{\sigma,j}}du_{j} \,\, 
P_{A, B}(u_{\sigma(1)}, \ldots , u_{\sigma(n)}; 
\lambda_{1}, \ldots , \lambda_{n}) \nn \\ 
&& \qquad {}\times \prod_{j=1}^{n-1}\left( \frac{e^{\pi i(n-2j)u_{j}}}
{\prod_{k=2}^{n}\sinh{\pi i(u_{j}-\lambda_{k})}} \right)
\frac{e^{\pi i(-n+1)u_{n}}}{\prod_{k=1}^{n}\sinh{\pi i(u_{n}-\lambda_{k})}}.
\nn
\end{eqnarray}
Here the contour $C_{\sigma,j}$ is determined by 
\be
C_{\sigma , j}=\left\{ 
\begin{array}{ll} 
C^{+} & \hbox{if}\,\, 1\le \sigma^{-1}(j) \le r, \\ 
C^{-} & \hbox{if} \,\, r+1 \le \sigma^{-1}(j) \le n. 
\end{array} \right. 
\en 
Note that the factor 
$e^{\pi i(n-1)\lambda_{1}}\prod_{j<k}\sinh{\pi i(\lambda_{j}-\lambda_{k})}$ 
converges when $\mu \to +\infty$. 
In the following we prove that 
each integral in the sum of the right hand side of \eqref{eq:asymp-decomp} 
converges in the limit as $\mu \to +\infty$. 

Decompose the contour for $u_{n}$ into two parts 
$C_{\sigma, n}=\Gamma+C'$, where 
$\Gamma$ is the segment $\Gamma^{+}$ or $\Gamma^{-}$ 
contained in $C_{\sigma, n}$, and  
$C'=C_{\sigma, n}\setminus \Gamma$. 
Simultaneously we decompose the integral 
\be 
\prod_{j=1}^{n}\int_{C_{\sigma, j}}du_{j}=
\prod_{j=1}^{n-1}\int_{C_{\sigma, j}}du_{j}\int_{\Gamma}du_{n}+
\prod_{j=1}^{n-1}\int_{C_{\sigma, j}}du_{j}\int_{C'}du_{n}. 
\en 
Further we decompose the integration in the first term over 
$C_{\sigma, 1} \times \cdots \times C_{\sigma, n-1}$ 
into the following parts: 
\be 
&& 
D_{1}=\{(u_{1}, \ldots , u_{n-1}) \in 
C_{\sigma, 1} \times \cdots \times C_{\sigma, n-1} | 
-\mu/3 \le {\rm Im}u_{j} \le \mu/3 \}, \\ 
&& 
D_{2}=(C_{\sigma, 1} \times \cdots \times C_{\sigma, n-1})\setminus D_{1}. 
\en 
Thus the integral in \eqref{eq:asymp-decomp} is decomposed as 
\bea 
\int_{D_{1}}\prod_{j=1}^{n-1}du_{j}\int_{\Gamma}du_{n}+
\int_{D_{2}}\prod_{j=1}^{n-1}du_{j}\int_{\Gamma}du_{n}+ 
\prod_{j=1}^{n-1}\int_{C_{\sigma, j}}du_{j}\int_{C'}du_{n}. 
\label{eq:int-decomp} 
\ena 
We consider the limit of these three parts separately. 

Let us consider the first integral of \eqref{eq:int-decomp}. 
Change the variable $u_{n} \mapsto u_{n}+\lambda_{1}$. 
Then the integral is equal to 
\bea 
&& 
\int_{D_{1}}\prod_{j=1}^{n-1}du_{j} 
\int_{\pm \epsilon-\frac{\mu}{3}i}^{\pm \epsilon+\frac{\mu}{3}i}du_{n} 
P_{A, B}(\ldots , u_{n}+\lambda_{1}, \ldots ; \lambda_{1}, \ldots , \lambda_{n}) 
\label{eq:int-1} \\ 
&& \quad {}\times 
\prod_{j=1}^{n-1}\left( 
\frac{e^{\pi i(n-2j)u_{j}}}
{\prod_{k=2}^{n}\sinh{\pi i(u_{j}-\lambda_{k})}} \right)
\frac{1}{\sinh{\pi i u_{n}}}
\prod_{k=2}^{n}
\frac{e^{-\pi i(u_{n}+\lambda_{1})}}{\sinh{\pi i(u_{n}+\lambda_{1}-\lambda_{k})}}, 
\nn
\ena 
where the sign $\pm \epsilon$ is $+$ or $-$ 
according to whether $\Gamma$=$\Gamma^{+}$ or $\Gamma^{-}$, 
respectively.

Note that 
\be 
P_{A, B}(\ldots , u_{n}+\lambda_{1}, \ldots ; \lambda_{1}, \ldots , \lambda_{n})=
O(1), \quad (\lambda_{1} \to \infty). 
\en 
Hence the integrand converges in the limit $\mu \to +\infty$ 
for fixed $u_{1}, \ldots , u_{n}$. 
To apply Lebesgue's convergence theorem, 
let us check that the integrand is bounded from above
by an integrable function. 
First we consider the rational part $P_{A, B}$. 
Recall that the contours $C^{\pm}, C^{\pm}-1$ and $C^{\pm}+1$ 
do not cross each other. 
Hence there exists a positive constant $d$ such that 
\bea 
|u_{j}-u_{k}\pm 1| \ge d 
\label{eq:denominator-bound} 
\ena 
for $j, k=1, \ldots , n-1$. 
For $(u_{1}, \ldots , u_{n-1}) \in D_{1}$ and 
$u_{n}=\pm \epsilon+iy, (-\mu/3 \le y \le \mu/3)$, we have 
\be 
|u_{j}-u_{n}-\lambda_{1}\pm 1| \ge 
|{\rm Im}(u_{j}-u_{n}-\lambda_{1})| \ge \mu/3 \quad 
(j=1, \ldots , n-1).  
\en 
Hence the rational part $P_{A, B}$ is upper bounded as 
\be 
&& 
|P_{A, B}
(\ldots , u_{n}+\lambda_{1}, \ldots ; \lambda_{1}, \ldots , \lambda_{n})| \\
&& {}\le 
\mu^{-2(n-1)}
Q_{1}(|u_{1}|, \ldots , |u_{n}|; |\lambda_{2}|, \ldots , |\lambda_{n}|; K; \mu), 
\en 
where $Q_{1}$ is a polynomial such that $\deg_{\mu}Q_{1}=2(n-1)$. 
Next we consider the trigonometric part. 
Note that the function $e^{\pi ix}/\sinh{\pi ix}$ 
is bounded in $\mathbb{C}\setminus U$, where $U$ is a union 
of a small open disk with a fixed radius around integer points.
Hence we have 
\bea 
&& 
\left| 
\prod_{j=1}^{n-1}\left( 
\frac{e^{\pi i(n-2j)u_{j}}}
{\prod_{k=2}^{n}\sinh{\pi i(u_{j}-\lambda_{k})}} \right)
\frac{1}{\sinh{\pi i u_{n}}}
\prod_{k=2}^{n}
\frac{e^{-\pi i(u_{n}+\lambda_{1})}}{\sinh{\pi i(u_{n}+\lambda_{1}-\lambda_{k})}} 
\right| 
\label{eq:trig-bound} \\ 
&& {}\le 
M \left| 
\prod_{j=1}^{n-1}\frac{1}{\sinh{\pi i (u_{j}-\lambda_{2})}} \, 
\frac{1}{\sinh{\pi i u_{n}}} \right| 
\nn 
\ena 
for some positive constant $M$.
The right hand side decreases exponentially as ${\rm Im}u_{j} \to \pm \infty$. 
Therefore we can apply Lebesgue's convergence theorem to \eqref{eq:int-1}. 

Next consider the second integral in \eqref{eq:int-decomp}. 
Change the variable $u_{n} \mapsto u_{n}+\lambda_{1}$. 
Then the integral is equal to \eqref{eq:int-1} 
with $D_{1}$ replaced  by $D_{2}$.
Let us prove that the integral vanishes in the limit $\mu \to +\infty$. 
To prove this we decompose $D_{2}$ as follows. 
Set 
\begin{eqnarray*}
U_{\pm}^{(k)}=\{\pm{\rm Im}u_{j}>\mu/3\} \cap D_{2}  
\end{eqnarray*}
for $k=1, \ldots , n-1$. 
Then $D_{2}=\cup_{k}(U_{+}^{(k)}\cup U_{-}^{(k)})$. 
We prove that the integral over each set 
$U_{\pm}^{(k)} \times \{ \pm \epsilon +iy | -\mu/3 \le y \le \mu/3\}$ 
vanishes in the limit. 
Here we consider the case of $k=1$. 
The argument is similar for the other cases. 
To prove the vanishing, we construct an upper bound for the integrand by 
a certain integrable function of the form 
\bea 
e^{-c \mu}
\sum_{k=-(n-1)}^{n-1}\mu^{k}
R_{k}(u_{1}, \ldots , u_{n}; \lambda_{2}, \ldots , \lambda_{n}), 
\label{eq:vanishing-bound} 
\ena 
where $c$ is a positive constant. 
First consider the rational part $P_{A, B}$. 
We have the inequality \eqref{eq:denominator-bound} and 
\be 
|u_{j}-u_{n}-\lambda_{1}\pm 1| \ge d
\en 
for $j=1, \ldots , n-1$. 
Hence we have an upper estimate of the form 
\be 
&& 
|P_{A, B}(\ldots , u_{n}+\lambda_{1}, \ldots ; \lambda_{1}, \ldots , \lambda_{n})| \\
&& {}\le 
\mu^{-(n-1)}
Q_{2}(|u_{1}|, \ldots , |u_{n}|; |\lambda_{2}|, \ldots , |\lambda_{n}|; K; \mu), 
\en
where $Q_{2}$ is a polynomial such that $\deg_{\mu}Q_{2}=2(n-1)$. 
Next consider the trigonometric part. 
We have the inequality \eqref{eq:trig-bound}. 
Apply the following inequality to the factor
$1/\sinh{\pi i (u_{1}-\lambda_{2})}$: 
\begin{eqnarray*}
\left|\frac{1}{\sinh{\pi i u}} \right| \le 
\frac{2e^{-\pi |{\rm Im}\,u|}}{1-e^{-\pi \mu/3}}
\le 
\frac{2e^{-\pi \mu/12}\cdot e^{-\pi|{\rm Im}\, u|/3}}{1-e^{-\pi \mu/3}}
 \quad (\pm {\rm Im}u >\mu/6 >0). 
\label{eq:key-estimate} 
\end{eqnarray*}
Thus we get the upper bound of the form \eqref{eq:vanishing-bound}. 

Finally let us consider the third integral in \eqref{eq:int-decomp}.  
It also vanishes in the limit $\mu \to +\infty$ as follows. 
The integrand is given in \eqref{eq:asymp-decomp}. 
By the same argument above we have 
an upper bound for the rational part $P_{A, B}$ 
of the form 
\be 
\mu^{-(n-1)}
Q_{3}(|u_{1}|, \ldots , |u_{n}|; |\lambda_{2}|, \ldots , |\lambda_{n}|; K; \mu), 
\en 
where $Q_{3}$ is a polynomial such that $\deg_{\mu}Q_{3}=n-1$. 
The trigonometric part is estimated as follows: 
\be 
&& 
\left| 
\prod_{j=1}^{n-1}\left( 
\frac{e^{\pi i(n-2j)u_{j}}}
{\prod_{k=2}^{n}\sinh{\pi i(u_{j}-\lambda_{k})}} \right)
\frac{e^{\pi i(-n+1)u_{n}}}{\prod_{k=1}^{n}\sinh{\pi i(u_{n}-\lambda_{k})}} 
\right| \\ 
&& {}\le 
M 
\left| 
\prod_{j=1}^{n-1}\frac{1}{\sinh{\pi i (u_{j}-\lambda_{2})}} \, 
\frac{1}{\sinh{\pi i(u_{n}-\lambda_{1})}} \right|. 
\en 
Now apply \eqref{eq:key-estimate} to the factor 
$1/\sinh{\pi i(u_{n}-\lambda_{1})}$, 
and we can see the vanishing in the limit $\mu \to +\infty$. 

{}From the consideration above we find that $h_{n}$ 
converges in the limit \eqref{eq:ang-domain}. 
It remains to show the limit is equal to 
the right hand side of \eqref{eq:reg_at_infty}. 
Denote the limit by $\hat{h}_{n}(\lambda_{2}, \ldots , \lambda_{n})$. 
{}From the equation \eqref{eq:rqKZ}, using
\[
{\rm lim}_{\lambda_1\rightarrow\infty}A_{\bar1}(\lambda_1,\ldots,\lambda_n)=-1,
\]
we obtain
\begin{eqnarray*}
\hat{h}_{n}(\lambda_{2}, \ldots , \lambda_{n})_{1, 2, \cdots , n,
 \bar{n}, \cdots , \bar{2}, \bar{1}}=
{}-\hat{h}_{n}(\lambda_{2}, \ldots , \lambda_{n})_{\bar{1}, 2, \cdots , n,
 \bar{n}, \cdots , \bar{2}, 1}. 
\end{eqnarray*}
Namely, the limit $\hat h_n$ is a singlet in the space $V_1\otimes V_{\bar1}$.
%
{}From this and \eqref{eq:n_to_n-1}, we get \eqref{eq:reg_at_infty}.

\medskip
{\it Acknowledgments.}\quad
HB and FS are grateful to V.E. Korepin for previous collaboration and 
many useful discussions. HB would also like to thank F.Goehmann, A.Klumper, 
M.Shiroishi and M.Takahashi for useful discussions
as well as ISSP of Tokyo University where work on
this paper was partially done.

Research of HB was supported by INTAS grant \#00-00561 and
by the RFFI grant \#04-01-00352.
Research of MJ was partially supported by 
the Grant-in-Aid for Scientific Research B2--16340033.
Research of TM was partially supported by 
the Grant-in-Aid for Scientific Research A1--13304010.
Research of FS was supported by INTAS grant \#00-00055
and by EC network  "EUCLID",
contract number HPRN-CT-2002-00325.

This work was started during the workshop, 21COE RIMS Research Project 2004,
Quantum Integrable Systems and Infinite Dimensional Algebras,
February 4-24, 2004.

\end{document}